\documentclass{mathincs}
\usepackage{bbm}
\usepackage{algpseudocode}
\usepackage{algorithm}
\usepackage{algorithmicx}
\usepackage{caption}
\usepackage{subfigure}
\usepackage{mathrsfs}
\usepackage{soul}
\usepackage{tikz}
\newcommand{\ph}{\varphi}

\def\mR{\mathbb{R}}

\def\tx{{\tilde x}}

\def\definedas{\stackrel{\Delta}{=}}

\newcommand{\cX}{{\mathcal X}}

\newcommand{\cN}{{\mathcal N}}

\newcommand{\cL}{{\mathcal L}}

\newcommand{\cZ}{{\mathcal Z}}

\def\tZ{{\tilde Z}}

\def\real{\mR}
\def\bc{\begin{center}}
\def\ec{\end{center}}

\newcommand{\beq}{\begin{eqnarray}}
\newcommand{\eeq}{\end{eqnarray}}
\newcommand{\beqq}{\begin{eqnarray*}}
\newcommand{\eeqq}{\end{eqnarray*}}
 \theoremstyle{definition}
 
 \theoremstyle{remark}

 \numberwithin{equation}{section}

\begin{document}
%
%
%
\title{Modeling of Persistent Homology}
\author{Sarit Agami}
\address{
Andrew and Erna Viterbi Faculty of  Electrical Engineering\\
Technion -- Israel Institute of Technology}
\email{sarit.agami@mail.huji.ac.il}
\thanks{Research supported in part by  URSAT, ERC  Advanced Grant 320422.}

\author{Robert J.\ Adler}

\address{
Andrew and Erna Viterbi Faculty of  Electrical Engineering\\
Technion -- Israel Institute of Technology}

\email{radler@technion.ac.il}


\date{August 8, 2017}

\begin{abstract}
Topological Data Analysis (TDA) is a novel statistical technique,  particularly  powerful  for the analysis of large and high dimensional data sets. Much of TDA is based on the tool of  persistent homology,  represented visually via  persistence diagrams. In an earlier paper  we  proposed a parametric representation for the probability distributions of  persistence diagrams, and based on it provided a method for their replication.  Since the typical situation for  big data is that only  one  persistence diagram is  available, these replications allow for conventional statistical inference, which, by its very nature, requires some form of replication. In the current paper we continue this analysis,  and further develop its practical statistical methodology, by investigating a wider class of examples  than treated previously.
\end{abstract}

\maketitle

\section{Introduction and setting}
The notion of persistent homology arises when one has a filtration of spaces; viz.\ a sequence (or continuum) of spaces
$\cZ_1\subseteq\cZ_2\subset\dots$ (or $\cZ_t$, with $\cZ_s\subseteq\cZ_t$ whenever $s\leq t$) and one is interested in how homology changes as one moves along the sequence. For a typical example, suppose that $\cZ$ is a nice space, and let  $f:\cZ\to\real$ be a smooth function. Denote by $\cZ_u$ the filtration of excursion, or upper-level, sets
\beq
\label{Au-defn}
\cZ_u \ \definedas \{z\in \cZ : f(z) \in [u,\infty)\} \
\equiv\ f^{-1}([u,\infty)).
\eeq

A useful way to describe persistent homology is via the notion of barcodes.
Assuming that $\dim (\cZ)=D$,  the smoothness of $f$ implies
that, if $\cZ_u$ is non-empty, then $\dim (\cZ_u)$ will typically also be $D$.  A barcode for the excursion sets of $f$ is then a collection of $D+1$ diagrams,
one for each collection of homology groups of common order. A bar in the
$k$-th diagram, starting at $u_1$ and
ending at $u_2$ ($u_1\geq u_2$) indicates the existence of a generator
of $H_k(\cZ_u)$ that appeared at level $u_1$ and disappeared at level $u_2$.
A different, and visually helpful, representation of a bar is as a point $(d,b)$ in the plane. Each bar has a `birth time' $b$ and `death time' $d$, where $d<b$ since, as described above, the filtration is for upper level sets, and we index these by levels descending from $+\infty$. The collection of points $(d,b)$ corresponding to all the bars is called the persistence diagram.

(We shall assume that the reader is familiar with these concepts. Recent excellent and quite different books and reviews by   Carlsson \cite{Carlsson-review,CarlssonReview},
Edelsbrunner and Harer \cite{EdelsShortCourse,EdelsHarerSurvey,EdelsHarerBook}, Zomorodian \cite{Afra},
Oudot \cite{Oudot} and Ghrist
\cite{ghrist2014elementary} not only give
give broad  expositions of  homology, but  also treat the much newer subject of persistent homology. (For a description of the history of persistent homology see the Introduction to \cite{EdelsHarerSurvey}.))

Persistence diagrams almost always arise  as topological summaries of some underlying phenomenon, and, having been constructed, are typically subject to some kind of analysis. This can be thought of as a path
\begin{equation}
\label{eq:to}
phenomenon \ \to\  persistence\ diagram \ \to \ analysis.
\end{equation}
The analysis can be of various forms.  Wasserman \cite{wasserman-review} gives a comprehensive and up to date  review on topological data analysis from the viewpoint of statistics, but there  are also non-statistical approaches, many of which involve summarizing  the diagram with either a low dimensional vector of numerical descriptors, a large dimensional vector, or a real valued function. Many of these approaches adopt  techniques such as
 principal component analysis and support vector machines to analyse the summary data.

 What is common to all these approaches, however, is the need for multiple instances of the persistence diagram, which in practical situations, is  typically
 not a trivial requirement. Although, in some scenarios, multiple observations of the `phenomenon' of \eqref{eq:to} may be available, it is more common that only one observation of the phenomenon  is available, and so only one diagram. In those cases, the standard method to effectively increase the number of instances is via resampling, either of the phenomenon or the diagram. Virtually all of the above approaches have  examples of this method.

In a previous paper  \cite{adleragamprat} we entered the diagram \eqref{eq:to} at the intermediate step, by suggesting a new approach to providing multiple instances of a persistence diagram when, perhaps, only one such original diagram is available. This was done  via probabilistic modeling of persistence diagrams.
We shall briefly recall the basic ideas of    \cite{adleragamprat} in  Sections 2 and 3 below, and then develop them, in terms of more sophisticated examples than were treated there, in Section 4.


\section{Parametric model}
\subsection{The basic setup}
 \indent As above, let $\cZ$ be a compact subset of $\mathbb R^D$, typically a sub-manifold or stratified sub-manifold, and suppose that we observe a sample $\tZ_n=\{Z_1,\dots,Z_n\}$ drawn from a distribution $P$ supported on $\cZ$. 
 Based on this sample, we define a  kernel density estimator, $\hat f_n$, given by
\beq
\hat f_n(p) \ = \  \frac{1}{n(\sqrt{2\pi}\eta)^D}  \sum_{i=1}^{n} e^{{-\|p-z_i\|^2}/{2\eta^2}},\qquad p\in\mathbb R^D,
\eeq
 where $\eta >0$ is a bandwidth parameter for the Gaussian kernel defining $\hat f_n$.

 Our interest is in the persistence diagram generated by the upper-level set filtration generated by $\hat f_n$; viz.\  by the sets of \eqref{Au-defn} as $u$ decreases from $+\infty$.
  The death and birth points in  the diagram are denoted by ${(d_i,b_i)}_{i=1}^{N_k}$, where $N_k$ is the number of points in the diagram for the homology of order $k$. Typically, we shall treat only one order at a time, and drop the subscript on $N_k$.
  Define a new set of $N$ points $\tx_N =\{x_i\}_{i=1}^N$, with $x^{(1)}_i=d_i$ and
 $x^{(2)}_i=b_i-d_i$. That is, $\tx_N$ is a set of $N$ points in $\cX=\real\times\real_+$.
This (invertible) transformation has the effect of moving the points in the original persistence diagram downwards, so that the diagonal line projects onto the horizontal axis, but still leaves a visually informative diagram, called  the projected persistence diagram, or PPD, in \cite{adleragamprat} . The first step towards the goal of a statistical analysis of the persistence diagram  is to develop a parametric, probabilistic model for $\tx_N$.

\subsection{The model}
\label{sec:Gibbs}
Following \cite{adleragamprat}, as a first step of building a model for $\tx_N$, consider the Gibbs distribution,
\beq
\ph_\Theta(\tx_N) = \frac{1}{Z_\Theta} \exp ( - H_\Theta (\tx_N)),
\label{eq:Gibbs}
\eeq
where $\Theta$ is a multi-dimensional parameter, $H_\Theta :\cX\to\real$ is a `Hamiltonian' that
describes the `energy' of $\tx_N$, and $Z_\Theta$ is the  normalizing `partition function' required to make $\ph_\Theta$ a probability density.
 The next step is to choose the Hamiltonian, which needs to take into account the spread of the $N$ points,  along with interactions between neighboring
 points, which we shall think of as belonging to clusters. Regarding the spread, define
\beqq
\sigma_H^2 =  \sum_{x\in\tx_N} \big(x^{(1)} - \bar x^{(1)}\big)^2,\ \ \
\sigma_V^2 = \sum_{x\in\tx_N} \big(x^{(2)} \big)^2,
\eeqq
where $\bar x^{(1)}=N^{-1} \sum_{i=1}^N x_i^{(1)}$.
Then $\sigma_H^2$ is  the (un-normalized) variance of the horizontal points, and $\sigma_V^2$ is the $L_2$ power of the vertical points (not centered because of the non-negativeness of $x^{(2)}$). As for the  local interaction, for  $x\in\cX$ and for $k\geq 1$ let
$x^{nn}(k) \in \cX$ be the $k$-th nearest neighbor of $x$, and set
\beqq
\mathcal L_{\delta,k}(\tx_N) =  \sum_{x\in\tx_N} \|x-x^{nn}(k) \| \mathbbm{1}_{\{\|x-x^{nn}(k) \|\leq \delta \}  }.
\eeqq
Then, as in \cite{adleragamprat}, we choose the Hamiltonian
\beq
\label{eq:hamiltonian}
H_{\delta,\Theta}^K(\tx_N)
= \theta_H \sigma^2_H
+\theta_V  \sigma^2_V
+\sum_{k=1}^K  \delta^{-2}\theta_k \mathcal L_{\delta,k}(\tx_N),
\eeq
where $\Theta=(\theta_H,\theta_V,\theta_1,\dots,\theta_K)$, and $K$ is the maximal cluster size. The inclusion  of the normalising  parameter  $\delta^{-2}$ allows for the $\theta_k$ to be interpreted as  energy densities, and improves the numerical stability of parameter estimation.

 There are a number of reasons for  this choice of
Hamiltonian, among them:
\begin{itemize}
\item[(i)] Cluster expansions of this form have been successfully  employed in Statistical Mechanics for the best part of a century as a basic approximation tool in the study of particle systems. More specifically, for the model to be rich enough for TDA,
 one needs to choose the Hamiltonian from a parameterised family that comes close to spanning all `reasonable' functions on PPDs. Since  \cite{GunnarRing} showed that the ring of algebraic functions on the space of PPDs is spanned by  a family of monomials
closely related to functions of the form  \eqref{eq:hamiltonian}, this choice of  Hamiltonian is an effective one for our setting.

\item[(ii)] These distributions are often used not as exact models for PPDs, but rather as a tool in a perturbative analysis. In these cases, the convenience of the models is more important than whether or not they provide a perfect fit to PPD data. For more details see \cite{adleragamprat}, SI  (Sec.\ 2.2).
\end{itemize}

Finally, $\delta$ is determined by
\beq
  \label{eq:delta}
 \delta =\frac{\delta^{*}}{N^{\alpha_{k,d}}}
 \max\left(  \max |x^{(1)}_i - x^{(1)}_j|,\    \max |x^{(2)}_i - x^{(2)}_j| \right),
  \eeq
where $\alpha_{0,d}=1/d$, $\alpha_{k,d}={k/(k+1)d}$, for $k\geq 1$, $d$ is the dimension of the data underlying the persistence diagram,
 and $\delta^{*} $ is a data independent tuning parameter. Although we typically optimize over $\delta^{*}$, it  can be taken to be $N^{-1/2}$, as a global default.

\subsection{Estimation and model specification}
Since there is no analytic form for $Z_\Theta$ and it is impossible to compute it numerically in any reasonable time, we cannot estimate $\Theta$ by  maximum likelihood. A standard way around this, adopted in \cite{adleragamprat}, is to use  pseudolikelihood  estimation \cite{Besag,chalmond}; viz.\ to maximize the pseudolikelihood
\beq
\label{eq:pseudo}
L^K_{\delta,\Theta}(\tx_N)  \definedas
\prod _{x\in\tx_N}  f_\Theta \left(x\big|    \cN_{\delta,K}(x) \right).
\eeq
Here
$\cN_{\delta,K}(x)$ denotes the collection of the $K$ nearest neighbours of $x$ in $\tx_N$ whose  distance from $x$ is no greater than $\delta$, and
 \beq
  f_\Theta\left(x\big|    \cN_{\delta,K}(x) \right)=\frac{
 \exp \left(-H^K_{\delta,\Theta}\left(x\big|   \cN_{\delta,K}(x) \right)\right)
  }{
\int _{\real}\int _{\real_+} \exp \left(-H^K_{\delta,\Theta}\left(z\big|   \cN_{\delta,K}(x) \right)\right)\,dz^{(1)}dz^{(2)},
}
\label{eq:conditionalham}
\eeq
with
\beqq
H^K_{\delta,\Theta}\left(x\big|   \cN_{\delta,K}(x)\right)=\theta_H\left[x^{(1)}-\bar x^{(1)}\right]^2+\theta_V(x^{(2)})^2 +\sum_{k=1}^K\delta^{-2} \theta_k\cL_{\delta,k}\left(\cN_{\delta,K}(x)\right).
\eeqq
Optimal values of $K$ can be chosen via standard,  automated, statistical procedures such as AIC, BIC, etc \ (cf.\ \cite{burnham}). However,  considerable experimentation, much of it reported in \cite{adleragamprat}, leads to the conclusion that it suffices  to take $K=2$ or $K=3$, so that the largest cluster size is 3 or 4.

\section{Replicating persistence diagrams}
\subsection{MCMC}
\label{Sec:MCMC}
Once the parametric distribution of the points of the persistence diagram is available, as  in the pseudolikelihood \eqref{eq:pseudo}, simulated replications of the  diagram can be generated using a standard Metropolis-Hastings MCMC algorithm \cite{RobertCasella,Handbook}.
Firstly, given a $\tx_N$,  define a `proposal distribution'
$q(\cdot |\tx_N)$ as the bivariate Gaussian density,  with mean vector and covariance matrix identical to the empirical mean and covariance of the points in $\tx_N$, but restricted to $\real\times\real_+$.
Next, for two points $x,x^*\in\real\times\real_+$  define an `acceptance probability', according to which $x\in\tx_N$ is replaced by $x^*$, leading to the updated PPD $\tx_N^*$, as
\beqq
\rho \left(x, x^*\right)
= \min \left\{1,\frac{f_\Theta\left(x^*| N_{\delta,K}(x)\right) \cdot q(x|\tx^*_N)}{f_\Theta \left(x|  N_{\delta,K}(x)\right)  \cdot q(x^*|\tx_N)}\right\}.
\eeqq

\noindent Then the algorithm is Algorithm \ref{MCMC:algorithm}.

\begin{algorithm}
\caption{MCMC step updating diagram for $\tx_N$}
\label{MCMC:algorithm}
\begin{algorithmic}[1]
\State  $k =0$
 \State $k \gets k+1$
\State Choose $x^*$ according to $q(\cdot | \tx_N)$
\State Compute $\rho(x_k,x^*)$
\State Choose $U$ a standard uniform variable on $[0,1]$

\If{$U<\rho(x_k,x^*)$} set $x_k =x^*$
\EndIf
\If{$k<N$}  go to Step 2
\EndIf
\end{algorithmic}
\end{algorithm}

 To obtain $M$ approximately independent PPD's, \cite{adleragamprat}
adopt a procedure dependent on  three parameters, $n_b$, $n_r$ and $n_R$, as follows.
 Starting with the original PPD, run the algorithm for a burn in period. Then,  starting with the final PPD from the burn in, run the algorithm a further $n_b$ times, choosing the last output of this block of $n_b$ iterations as the first simulated PPD. Repeat  this procedure $n_r$ times, each time starting  with the most recently simulated PPD; viz.\ the output of the previous block. Finally, replicate the entire procedure $n_R$ times, for a total of
 $n= n_r\times n_R$ simulated PPDs.
The optimal choice of $n_b$, $n_r$ and $n_R$ typically depends on the specific problem, and is discussed in \cite{adleragamprat} SI   (Sec.\ 2.1).
The burn in period is determined, empirically, via the evaluation of the distance between the MCMC simulations and the original persistence diagrams.
Recall that, for two diagrams $D_1$ and $D_2$, the Wasserstein $p$-distance, $W_p(D_1,D_2)$, $p>0$, is  defined  as
\beq
\label{eq:SIWass}
      W_p\left(D_1,D_2\right) \ =\  \inf_\gamma \big(
      \sum_{u\in D_1} \|u-\gamma (u)\|_\infty^p\big)^{1/p}
      \eeq
where $\gamma$ ranges over all matchings between the points of $D_1$ and $D_2$, the latter having been augmented by adding all points on the diagonal. In the limit case of $p=\infty$ the Wasserstein distance is known as the bottleneck distance, which is the length of the longest edge in the best matching.

The distances between the MCMC simulations and the original persistence diagrams are measured via  the bottleneck and Wasserstein distances as the MCMC progresses, and the value of the burn in period is chosen as the point at which the initial rapid growth of the distance functions ceases. Given the collection of $M$ simulated PPDs, each PPD is converted back to a regular persistence diagram with the mapping {$x\to ( x^{(1)},x^{(1)}+x^{(2)})=(d,b)$ }of its component points.

\subsection{Identification of topological signals}
In many situations, which include all the examples that we shall treat in this paper, the most prominent features of the persistence diagram are generally deemed most likely to represent true features of the underling space, rather than artifacts of sampling or noise. By `prominent' we mean  those points in the diagrams which are  furthest from the diagonal. These are typically  called `topological signals', while the points  closer to the diagonal are considered to be `topological noise'.

One of the key challenges in persistent homology is to separate the signal from the noise.
The replicated persistence diagrams can be used to identify the topological signals by providing information about statistical variation. As  an example, consider  the order statistics of the distances of the points of the persistence diagram to the diagonal. That is, given the points $(d_i,b_i)$ of the persistence diagram, the order statistics  are $T_j$, the $j-$th largest among the differences $|b_i-d_i|$, $j=1,...,N$.

Denote by $\hat{T}$ the value of the relevant statistic based on the true persistence diagram, and denote by $\hat{T}^{*} $ the value of the relevant statistic based on the simulated persistence diagram.
Then one can calculate an empirical  confidence interval (percentile bootstrap) $[c_1,c_2]$ and define a point on the persistence diagram to be a signal if $\hat{T}<\hat{c}_{1} $  or $\hat{T}>\hat{c}_{2} $.
Note that since here $T$ is non-negative, we generally only consider  one-sided confidence intervals $[0,\hat{T}+\hat{c}_{2} ]$. In addition, one can calculate a one-side $p$-value as $\frac{1}{M} \sum _{m=1}^{M}{\mathbbm{1}}_{\left(\hat{T}_{m}^{*} \ge \hat{T}\right)} $, where $\hat{T}_{m}^{*}$ is $\hat{T}^{*}$ of the $m$-th simulated persistence diagram.

\section{Examples}
We now turn to the three examples which make up the  new material of this paper. In each of these we show how to use the methodology described in the previous sections to identify the homology of spaces $\mathcal Z$, when all that is available is the persistence diagram generated by the upper level sets of a smoothed empirical density from a sample. The examples that we shall treat are those of a 2-sphere, a 2-torus and a collection of three concentric circles in the plane. Each of these will teach us something different about the  behaviour of our methodology in practice.


\subsection{The two dimensional sphere}

\subsubsection{The data and fitting the model}

\begin{figure}[h!]
    \centering
    \subfigure[]
    {
        \includegraphics[width=1.3in, height=1.3in]{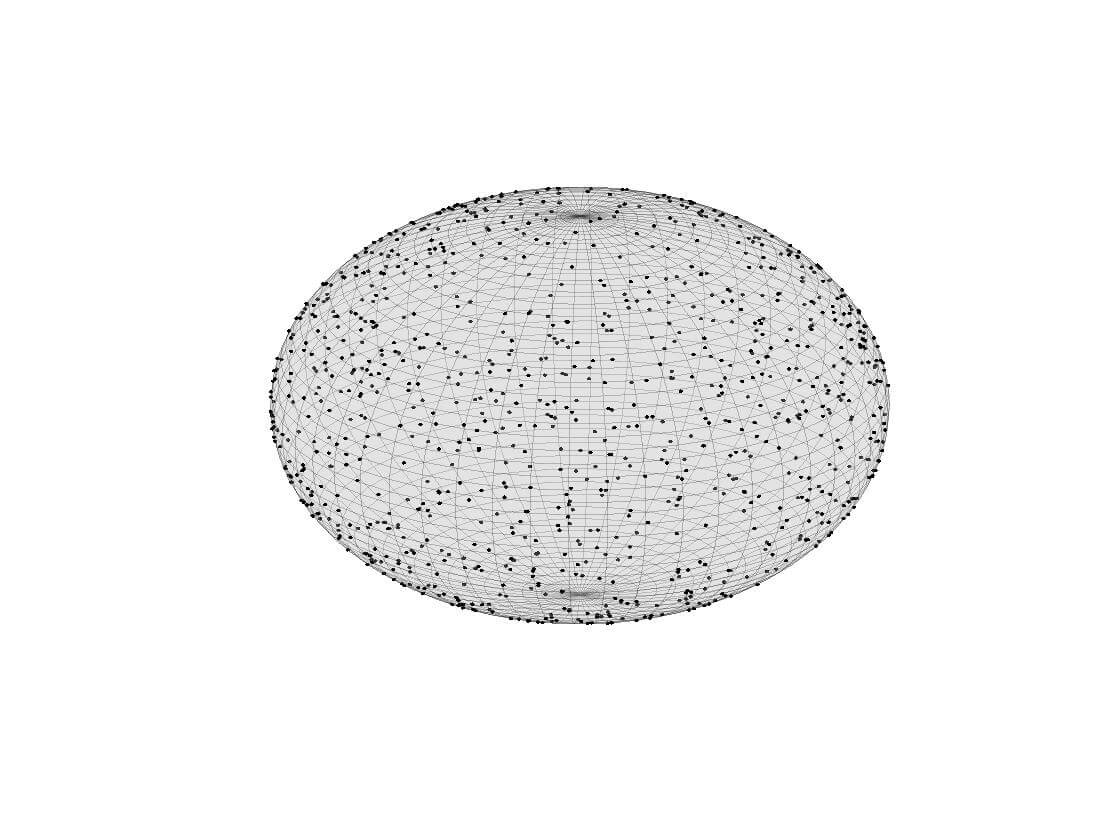}\ \  \ \
    }
       \subfigure[]
    {
        \includegraphics[width=1.3in, height=1.3in]{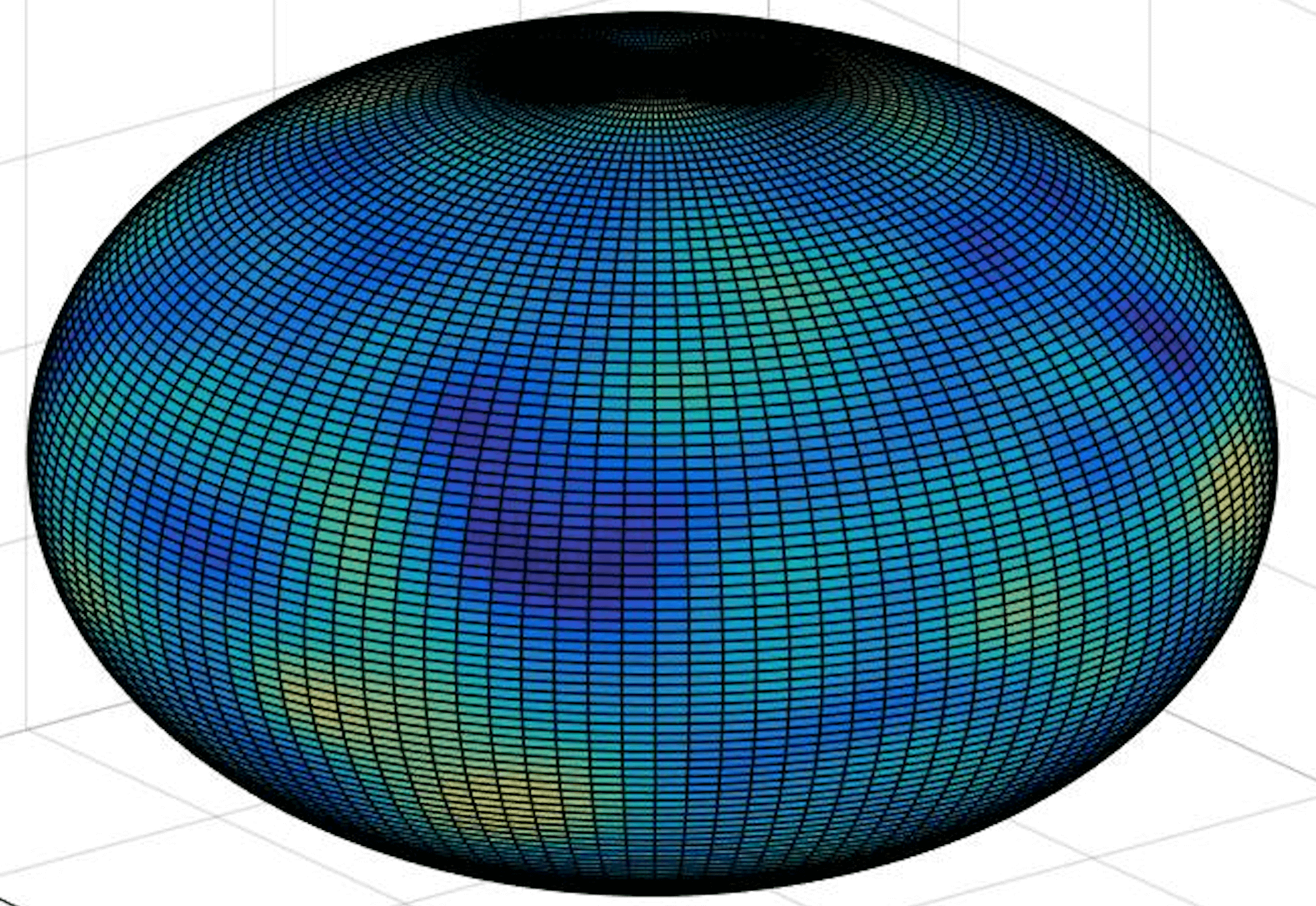} \ \ \ \
    }
       \subfigure[]
    {
        \includegraphics[width=1.3in, height=1.3in]{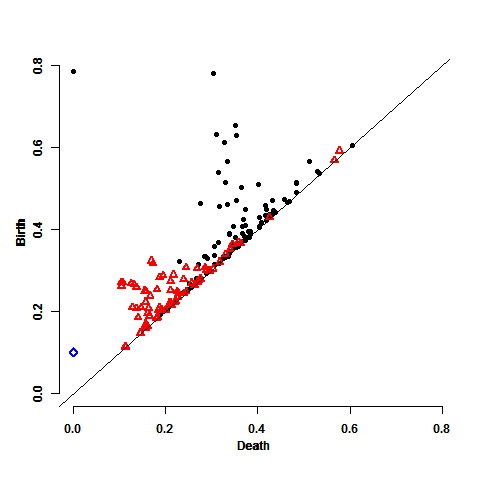}
    }
\caption{\footnotesize
 (a) Points sampled from a unit sphere. (b) The corresponding kernel density estimator, shown, for visual clarity, at  only a few quantized levels.  (c)  The corresponding persistence diagram for the upper level sets of the kernel density estimate on the full sphere. Black circles are  $H_0$ persistence points,  red triangles are  $H_1$ points, and the blue diamond is  the $H_2$ persistence point. Birth times are on the vertical axis.}
\label{fig:sphere}
\end{figure}

We start with a random sample of $n=1,000$ points from the uniform distribution on the sphere $S^2$ in $R^3$ with radius $r=1$, and smooth the data with a kernel density estimator of bandwidth of $\eta=0.1$. These are shown as Panels (a) and (b) of Figure \ref{fig:sphere}.

The corresponding persistence diagram of the upper level sets filtration of $\hat f_N$ is shown in Panel (c) of Figure  \ref{fig:sphere}. This diagram contains $N_0=110$ points corresponding to  the zeroth homology $H_0$, represented by  the black circles, $N_1=74$ points for the first homology $H_1$, represented by  the red triangles, and $N_2=1$ point for the second homology $H_2$, represented by the blue diamond. As described above, each point in the diagram is a `death-birth' pair $(d,b)$. Since we know that the upper level sets of $\hat f_N$ are characterized by having a single connected component and a single void, we expect to have one black circle somewhat isolated from the other points in the diagram and one blue diamond. The void does not have to be isolated from the other points due to a short lifetime of high dimensional homologies. This is in fact the case.

While the  persistence diagram in Figure  \ref{fig:sphere} performs as expected, and it is easy to identify the points that, a priori, we knew had to be there, there are many other points in the diagram which, were we not in the situation of knowing ahead of time, and we would have difficulty in knowing  how to discount.

 Note that there are more than  enough $H_0$ and $H_1$ points in Figure \ref{fig:sphere} to fit a spatial model to each of the two homologies.

Adopting the approach described in the first three sections of the paper, and working first with the $H_0$ persistence diagram without including the  `point at infinity\footnote{In all our persistence diagrams, the `point at infinity' is  the highest, leftmost point in the $H_0$ diagram. In essence, removing it from the analysis is much like working with reduced rather than standard homology, and has the effect of removing one generator from the $H_0$ diagram. Thus, in the statistical analysis to follow, it needs to be added, at the end, to all significant points found in the diagram.}', we estimated the parameters for a Gibbs distribution for the model with pseudolikelihood \eqref{eq:pseudo}, taking $K=3$. The estimate of $\delta$ was 0.0051. For this $\delta$, the estimates of $\Theta$ were $\theta _{1} = -0.0339$, $\theta _{2} = -0.0210$, $\theta _{3} = -0.0120$, $\theta _{H} =72.80$, and $\theta _{V} =39.50$.

In order to test how well the estimated  model  matches the persistence diagram, we followed the procedure described in  \cite{adleragamprat}. We generated 100 collections of samples from the 2-sphere according the same procedure that generated the original data, and for each  we fitted the  model  we found  for the original data set; viz.\  the model that includes the parameters $\theta _{1}$, $\theta _{2}$, $\theta _{3}$, $\theta _{H}$, and $\theta _{V}$.

\begin{figure}[h!]
    \centering
    \subfigure[]
    {
        \includegraphics[scale=.18]{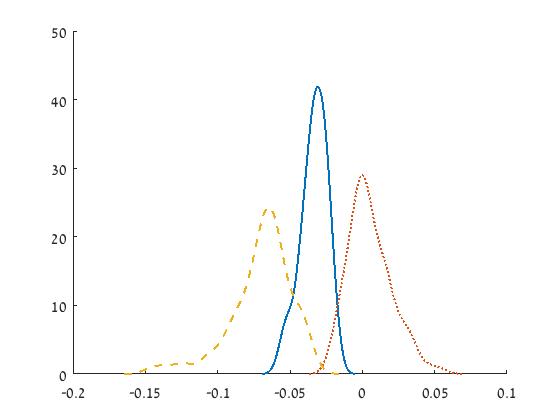}
    }
       \subfigure[]
    {
        \includegraphics[scale=.18]{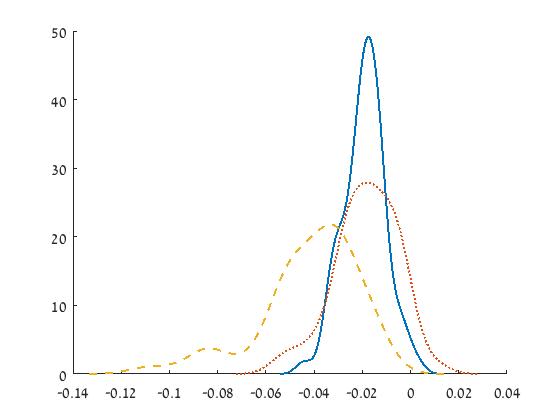}
    }
       \subfigure[]
    {
        \includegraphics[scale=.18]{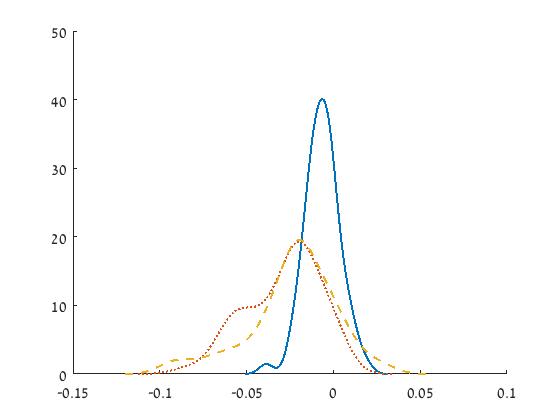}
    }
       \subfigure[]
    {
        \includegraphics[scale=.18]{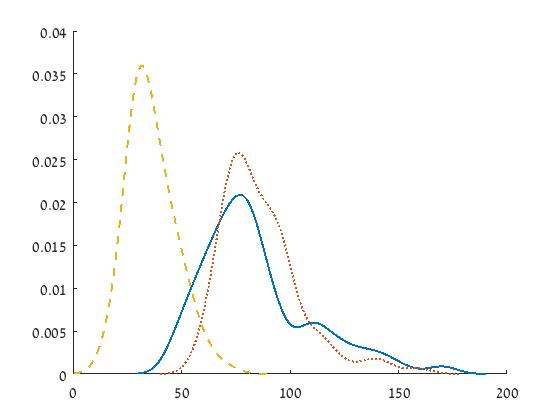}
    }
       \subfigure[]
    {
        \includegraphics[scale=.18]{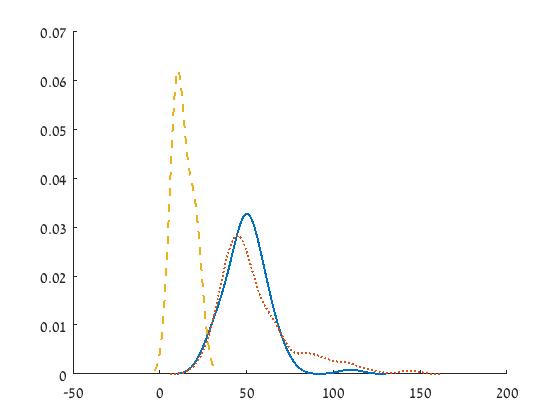}
    }

\caption{\footnotesize
 Smoothed empirical densities for the five parameter estimates in the Hamiltonian \eqref{eq:hamiltonian} for the
$H_0$ persistence diagram coming from the simulations of 2-sphere and from resampling, see text for details. (a) $\theta_1$, (b) $\theta_2$, (c) $\theta_3$, (d) $\theta_H$, (e) $\theta_V$.}.
\label{fig:resPDestsph1}
\end{figure}

The blue plot in Figure  \ref{fig:resPDestsph1} shows the (smoothed) empirical densities of the resulting  parameter estimates\footnote{In some of these simulations the sums  $L_{\delta,k}$ were identically zero for all $k=1,2,3$ simultaneously, since there were no $k$-th nearest neighbours at distance less than $\delta$. Consequently, the parameters $\theta_1$, $\theta_2$, and $\theta_3$ are all meaningless, and so these simulations (33 of them) were deleted from this part of the analysis. We shall do the same later on, in similar cases, without further comment.}.
  (We will discuss the other two plots only later, when considering different replication procedures.)
Overall, the results indicate that the estimation procedure  is stable, with an acceptable  spread.

\subsubsection{Replicating the persistence diagram}
\begin{figure}[h!]
    \centering
    \subfigure[]
    {
        \includegraphics[scale=.18]{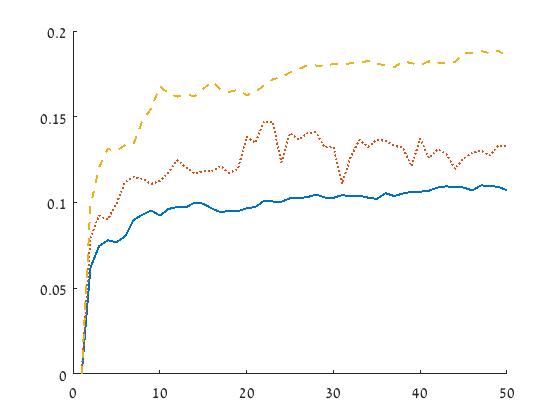}
    }
      \subfigure[]
    {
        \includegraphics[scale=.09]{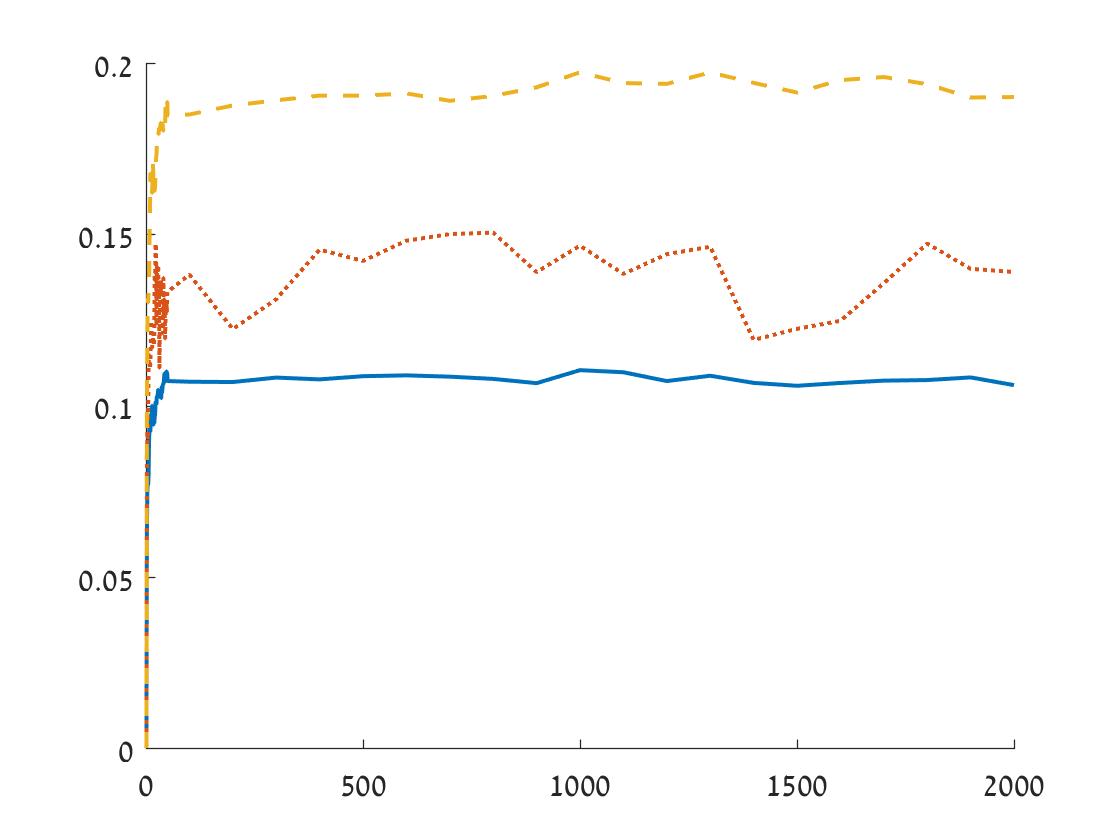}
    }
   \subfigure[]
    {
        \includegraphics[scale=.18]{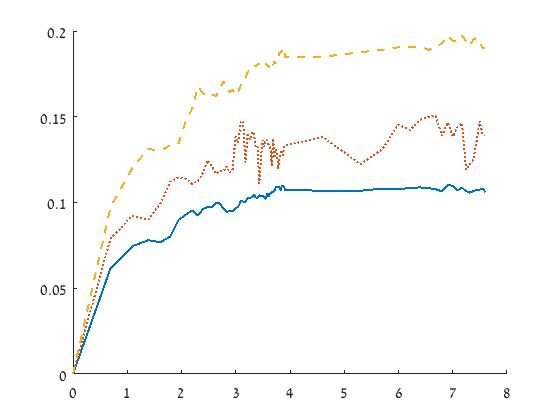}
    }
    \subfigure[]
    {
        \includegraphics[scale=.18]{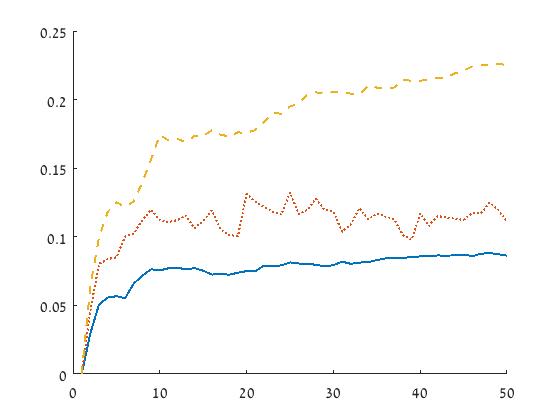}
    }
   \subfigure[]
    {
        \includegraphics[scale=.09]{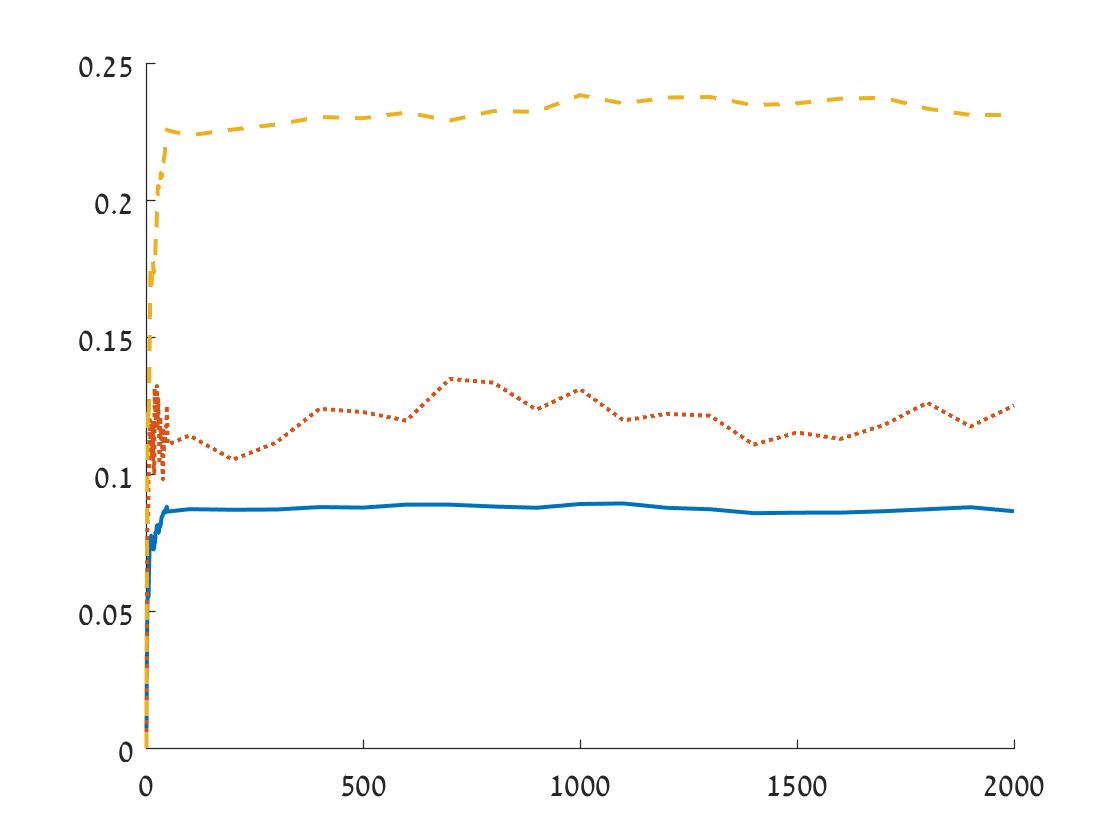}
    }
    \subfigure[]
    {
        \includegraphics[scale=.18]{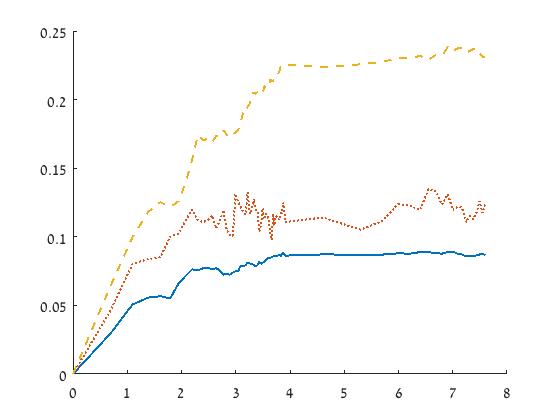}
    }
\caption{\footnotesize
    Growth of the bottleneck (a) and Wasserstein (d) differences of MCMC simulations from a specific persistence diagram (vertical axis), as a function of the number of steps $n_b$ (horizontal axis, $1\leq n_b\leq 50$)  averaged over 100 independent  persistence diagrams. Panels  (b) and  (e) take $1\leq n_b\leq 2,000$, while (c) and (f) show the same data but on a logarithmic scale.}.
\label{sphburn}
\end{figure}

For the calculation of the replicated persistence diagrams, we first need to determine the burn in period  which we shall use for  them. Following the procedure described in Section \ref{Sec:MCMC}
we calculated the bottleneck and  Wasserstein distances using the 100 simulated persistence diagrams of the previous subsection. The results are shown in  blue  in Figure  \ref{sphburn}.  

The first row in Figure  \ref{sphburn} shows the bottleneck distances, while the second row shows the $W_2$ differences. The first column shows the results of the first 50 steps of the MCMC algorithm on a linear scale. The second and third columns go out to 2,000 steps, first on a linear scale and then on a logarithmic scale. While the initial  growth of the  distances is rapid,  they eventually approach  their asymptotes at exponential rates. The rapidity is clear in Panels (a) and (d), and the exponential rate is clear from the linear behavior of the plot in  logarithmic scales. The point where the initial rapid growth of the distance functions ceases is  approximately 44 for the bottleneck distance and  47 in the  Wasserstein case. At 44 steps, therefore, the results of Figure  \ref{sphburn}
indicate that the dependence of the MCMC on the initial persistence diagram  has dropped significantly, while at the same time the MCMC has produced persistence diagrams remaining close to  the  true distribution.

\begin{figure}[h!]
    \centering
    \subfigure[]
    {
        \includegraphics[scale=.09]{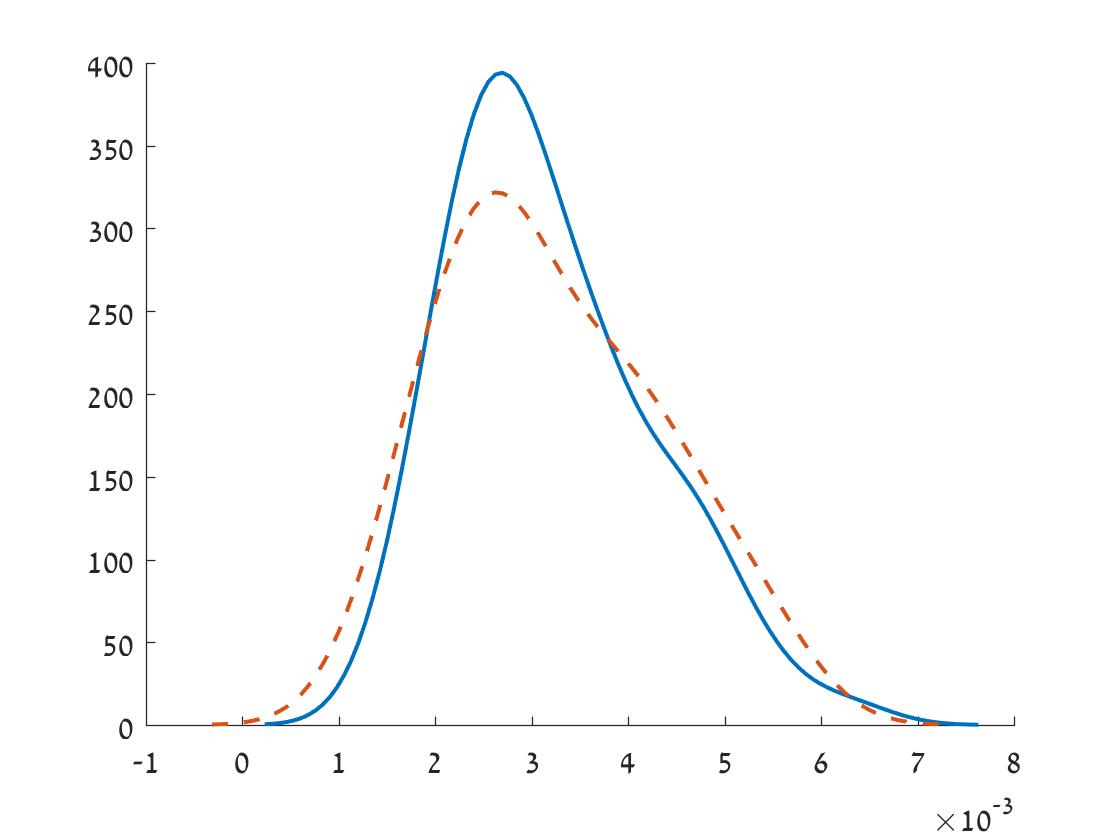}
    }
    \subfigure[]
    {
        \includegraphics[scale=.09]{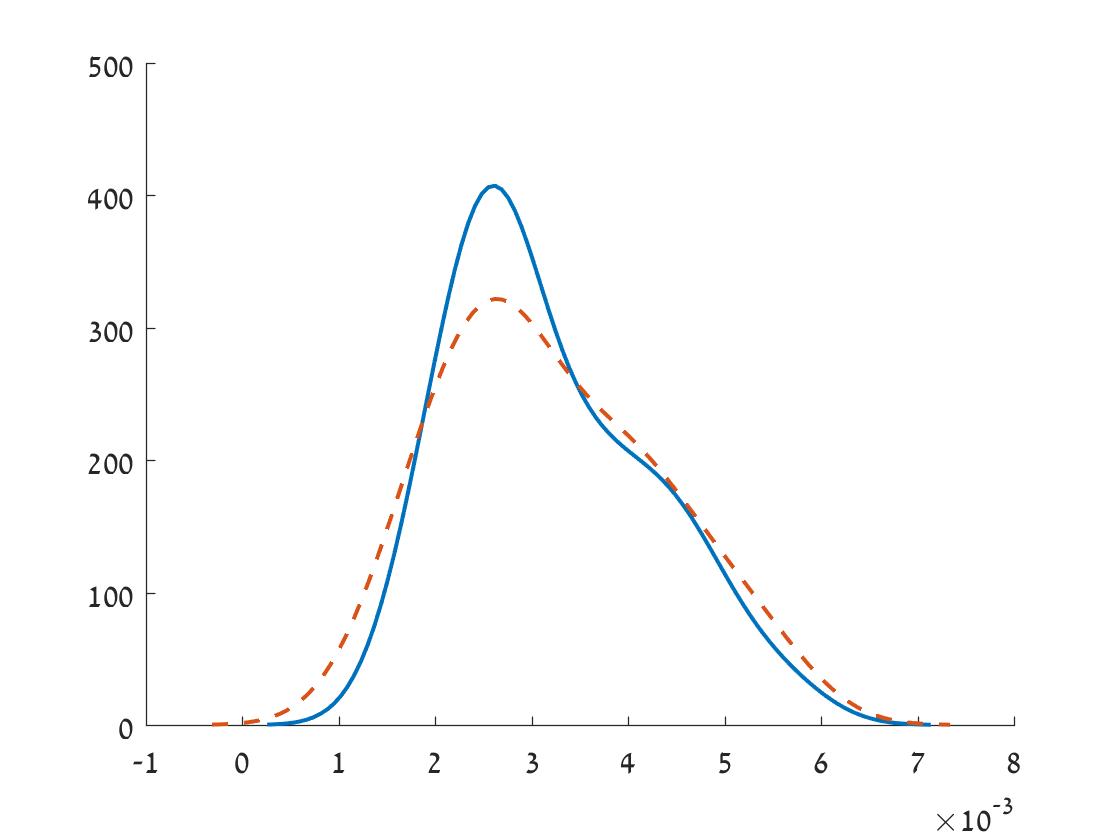}
    }
    \subfigure[]
    {
        \includegraphics[scale=.09]{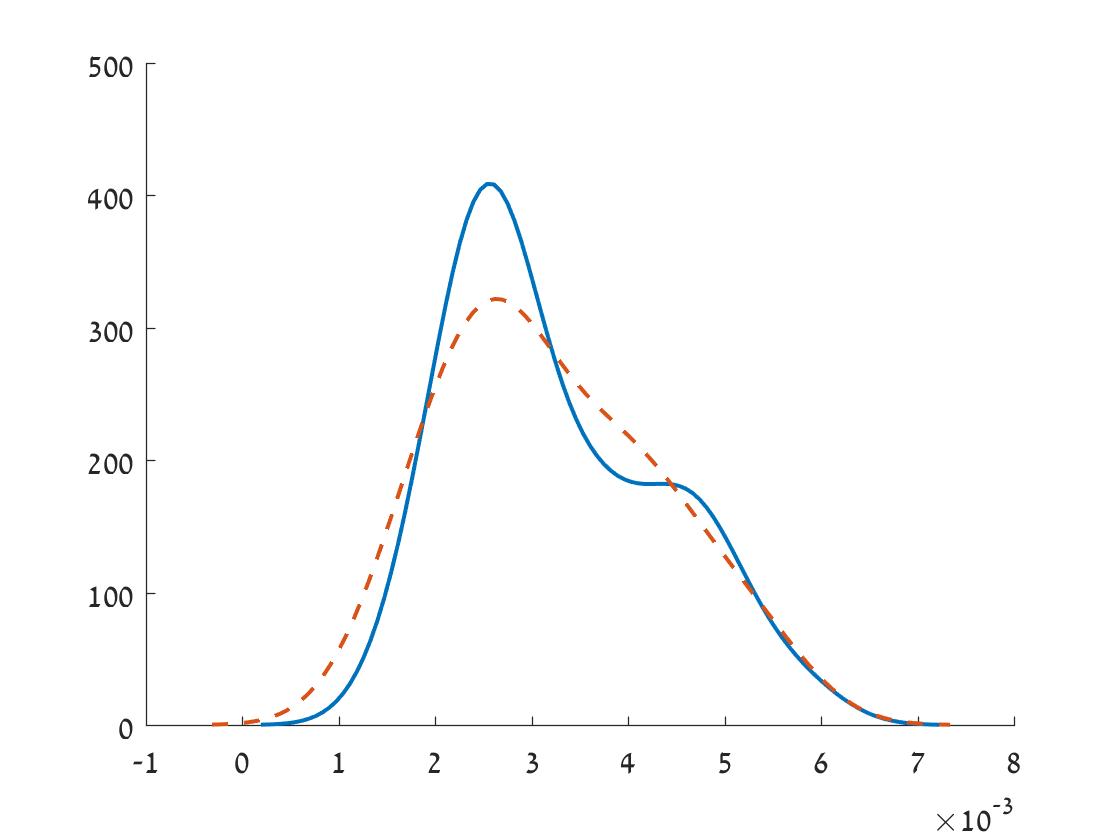}
    }
     \subfigure[]
    {
        \includegraphics[scale=.09]{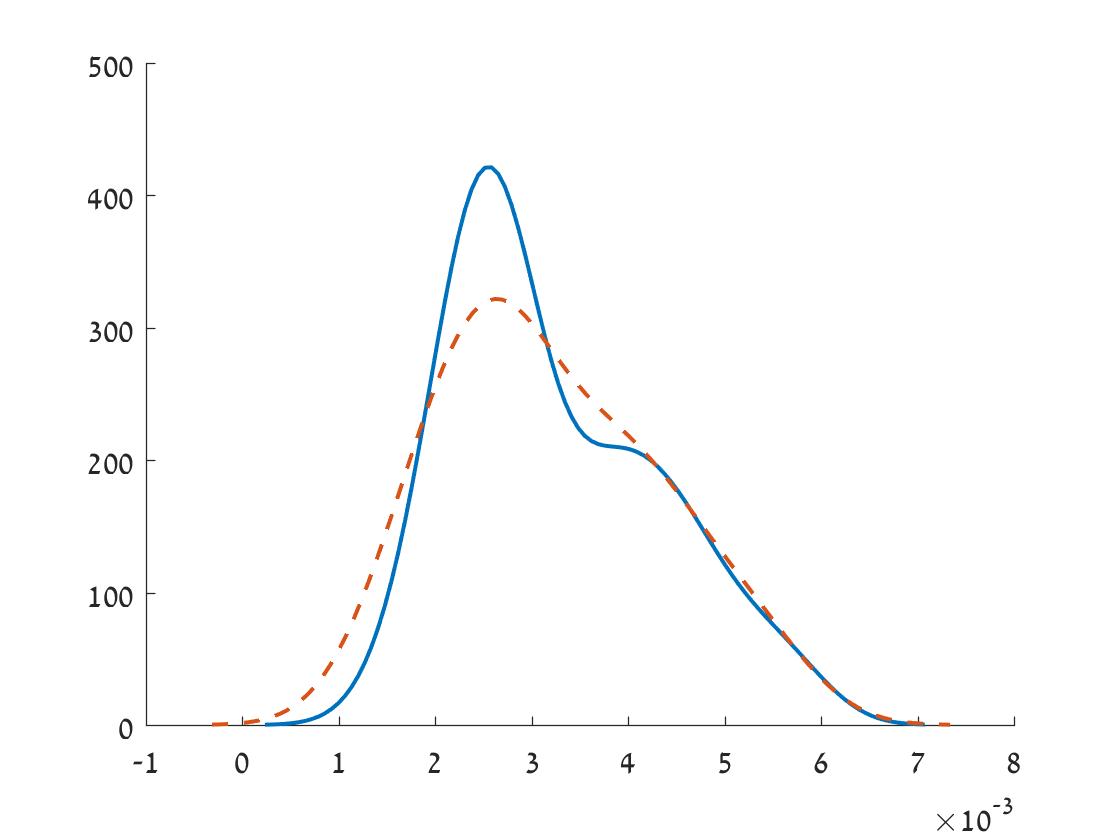}
    }
    \subfigure[]
    {
        \includegraphics[scale=.09]{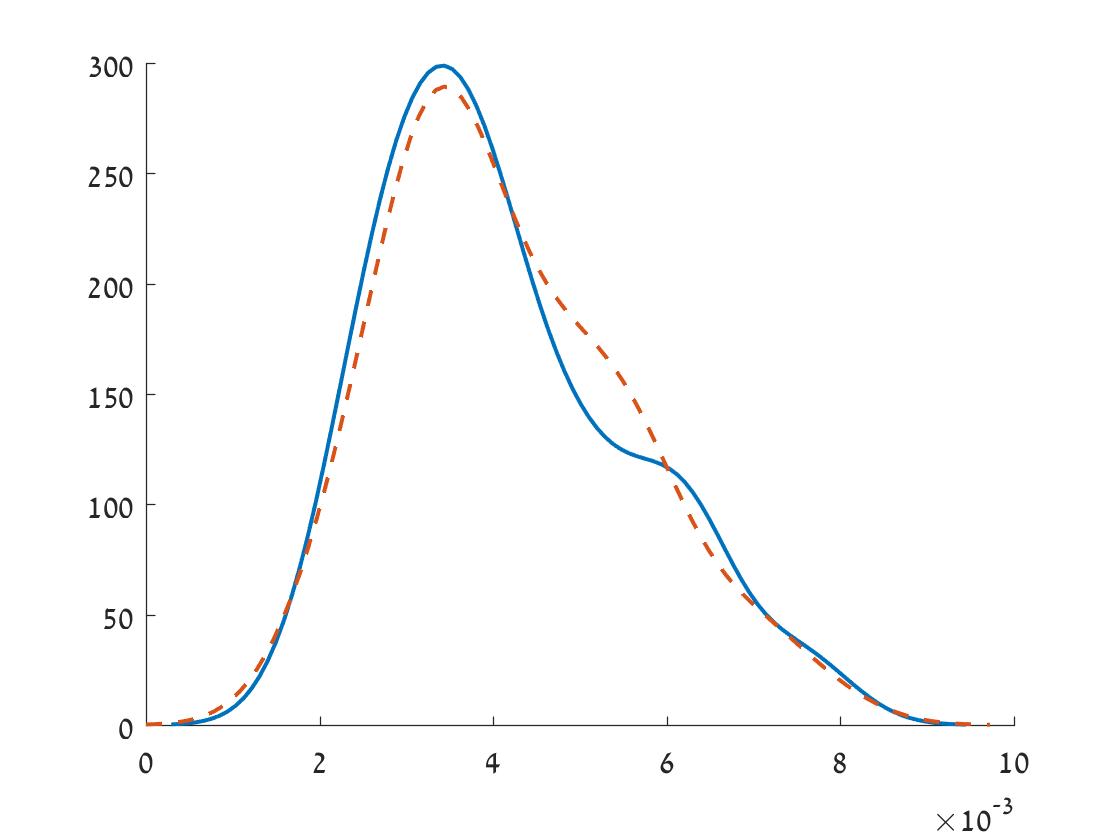}
    }
        \subfigure[]
    {
        \includegraphics[scale=.09]{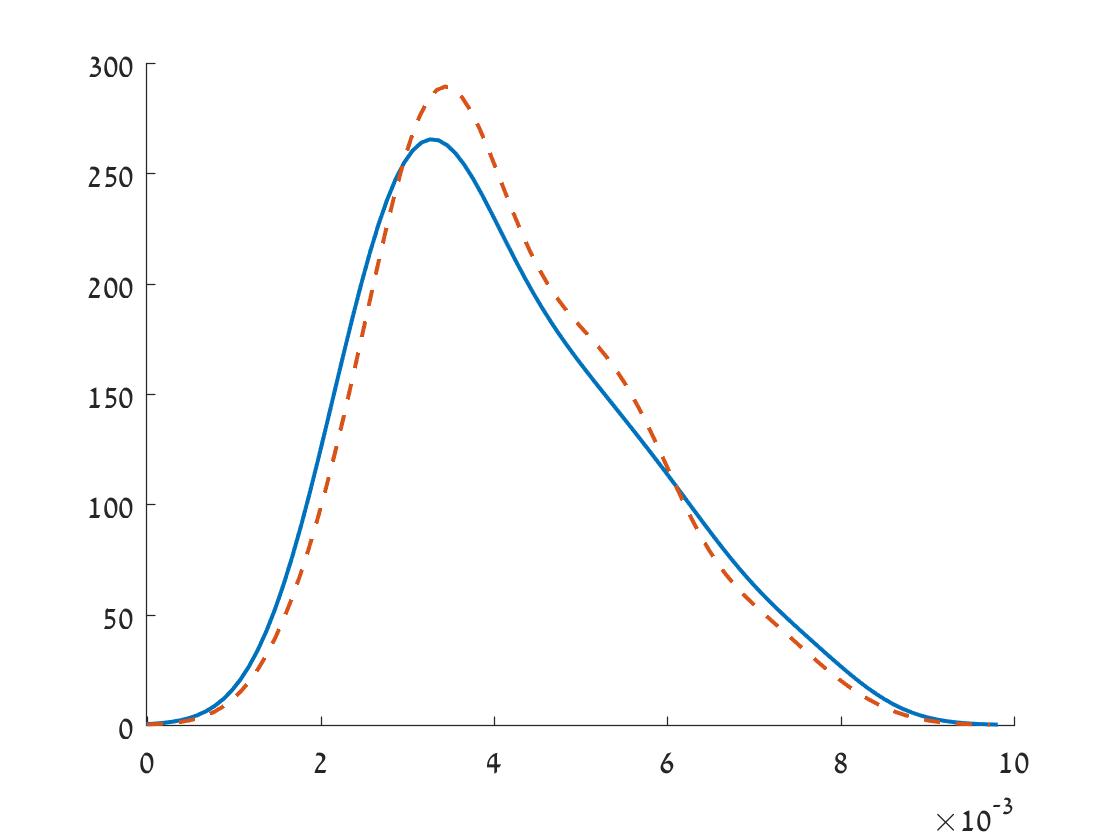}
    }
        \subfigure[]
    {
        \includegraphics[scale=.09]{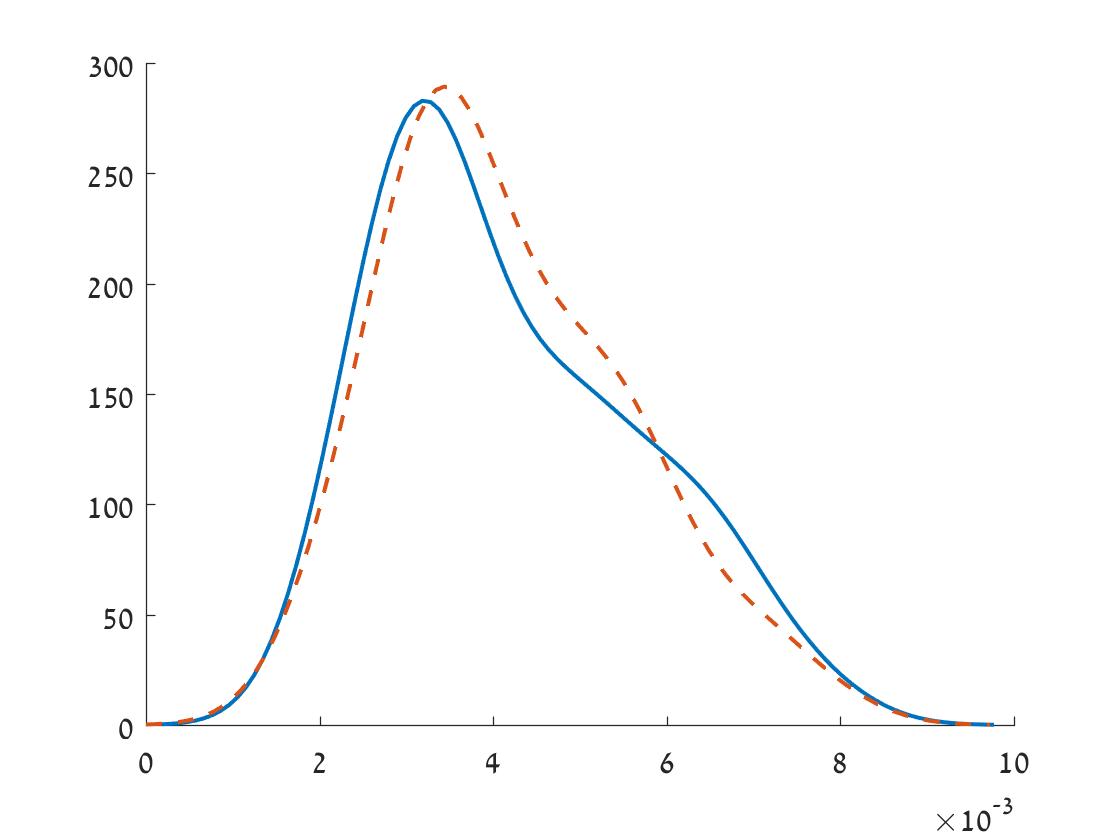}
    }
        \subfigure[]
    {
        \includegraphics[scale=.09]{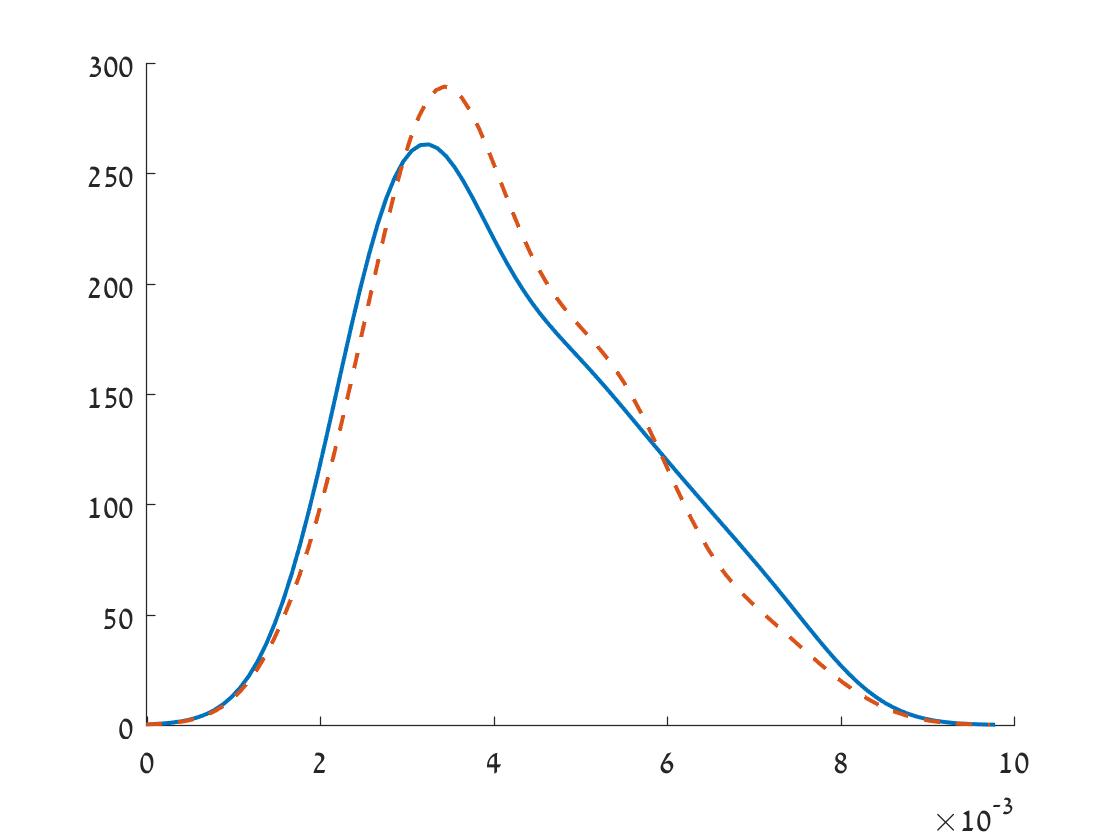}
    }
        \subfigure[]
    {
        \includegraphics[scale=.09]{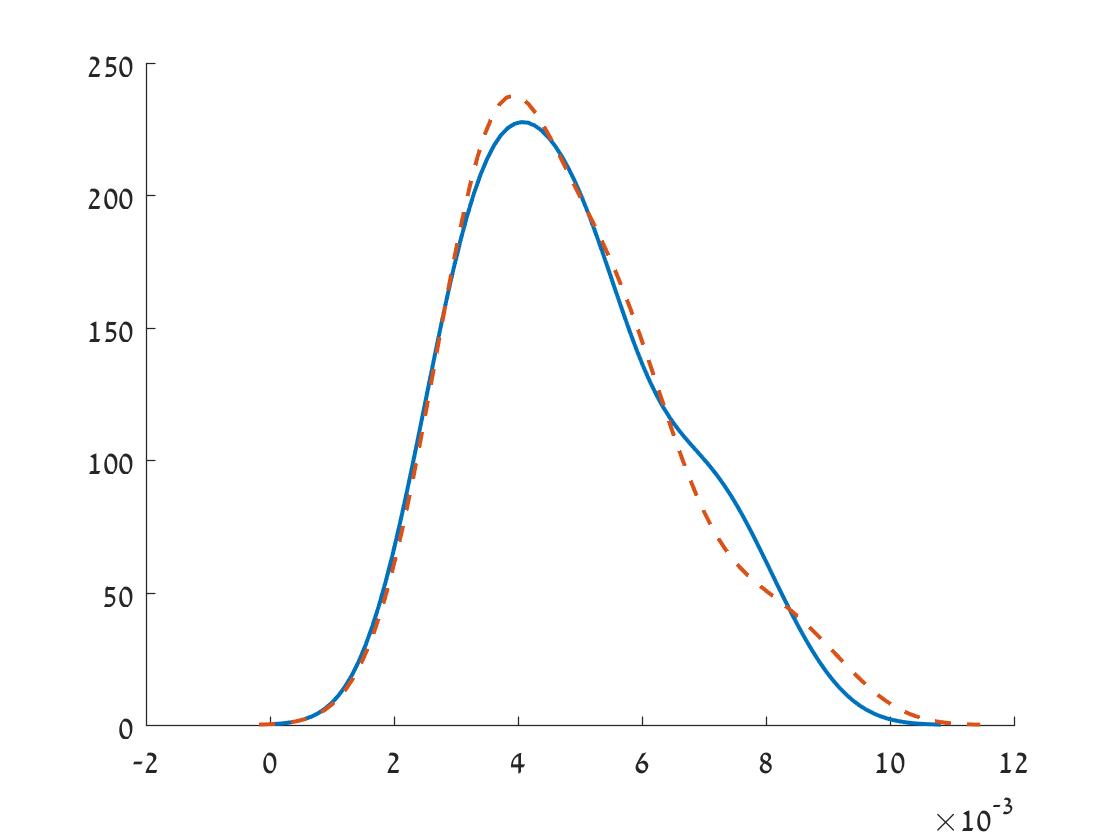}
    }
        \subfigure[]
    {
        \includegraphics[scale=.09]{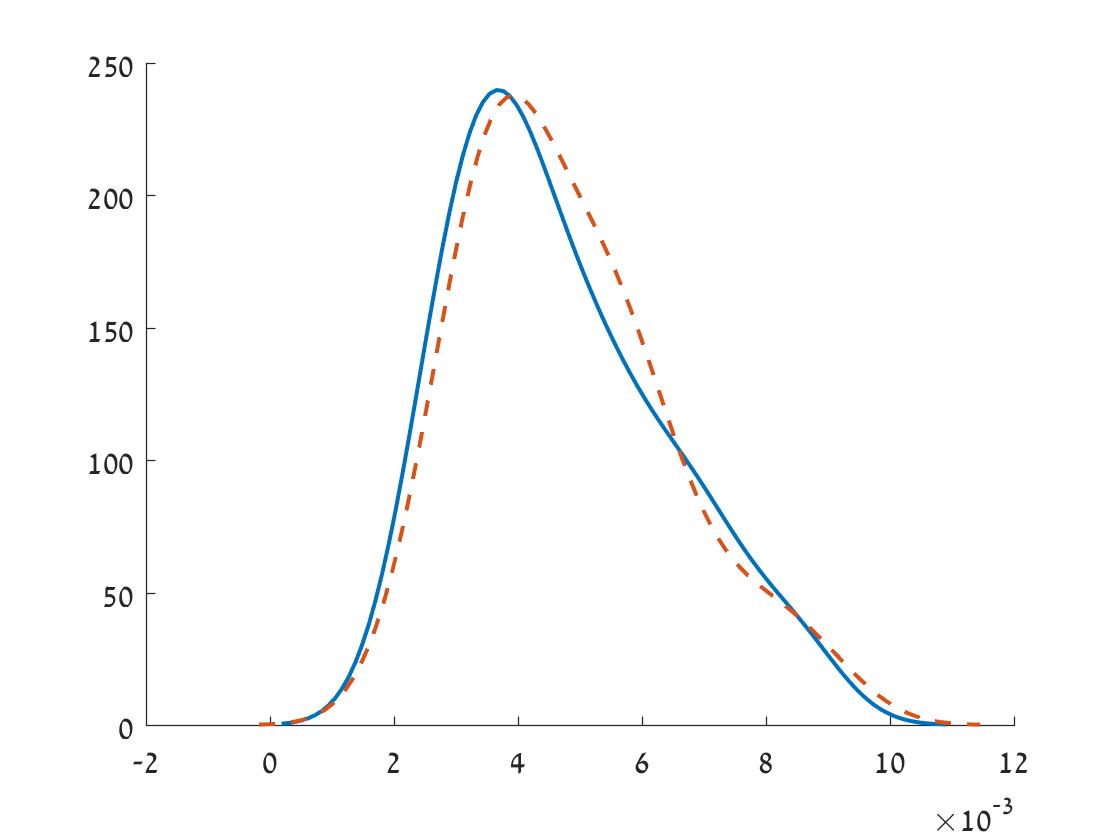}
    }
        \subfigure[]
    {
        \includegraphics[scale=.09]{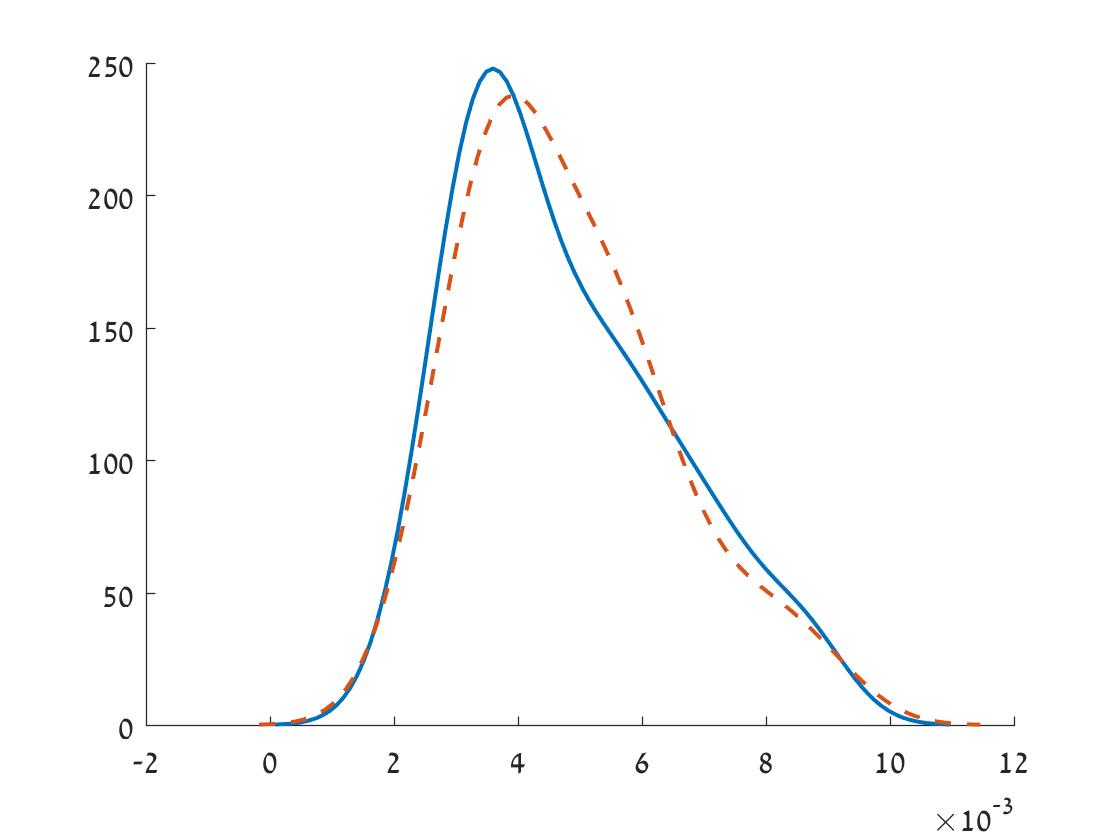}
    }
        \subfigure[]
    {
        \includegraphics[scale=.09]{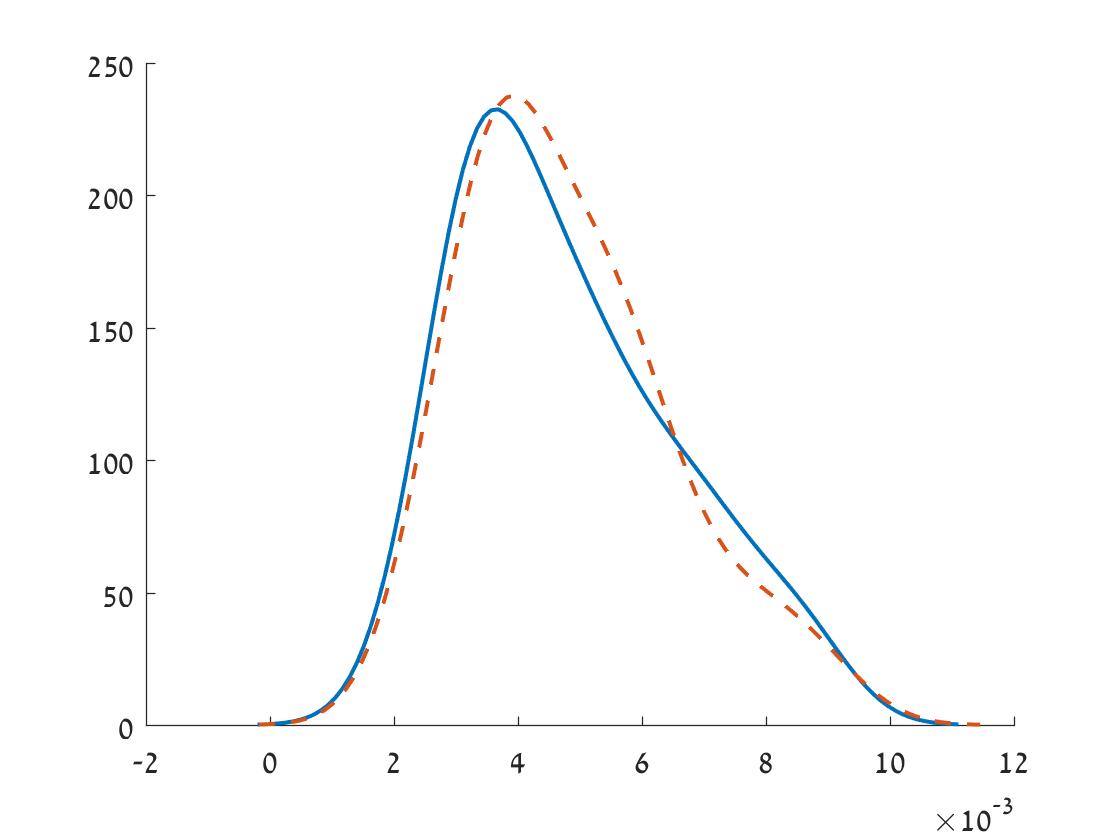}
    }
\caption{\footnotesize
    Summary statistics of average interaction strengths for 100 persistence diagrams. From left to right: cluster sizes 2, 3, and 4. From top to bottom,   after 10, 25, 50, and 1,000 MCMC steps. See text for details.}
\label{fig:mcmcsphr}
\end{figure}

In addition we considered summary statistics of the 100 persistence diagrams as the MCMC progressed, to see how well the simulations  replicate the statistical properties of the persistence diagrams. The results are presented in Figure \ref{fig:mcmcsphr}. Overall, the best fits are at  burn in of 10, 25 and 50, which is consistent with the results of Figure  \ref{sphburn}.

\subsubsection{Resampling}
The  replicated persistence diagrams described in the previous subsection were based on knowing, a priori, that the original data was generated by sampling from   a 2-sphere. The typical real-life situation is that one does not know the space generating the persistence diagram. (If one did, it would hardly be necessary to estimate its homology by sampling.) Consequently, we now look at resampling as a method for generating replications of the persistence diagram.

 There are two natural approaches based on resampling. One  is to  resample  from the original persistence diagram (``Setting I"), and another is to resample   from the original data (``Setting II"). We examined both these alternatives, repeating them 100 times.

The results of these approaches are the other two plots  of Figure  \ref{fig:resPDestsph1}.  The red (dot dashed) plot is   the smoothed empirical density for the parameter estimates based on resampled sets from the original persistence diagram, and the yellow (dashed) plot corresponds to  the resampled sets from the original data.


In order to assess the fit of the simulated data to the original, we computed, as previously, the bottleneck and the Wasserstein distances between the MCMC simulations and the data itself. The results are  in Figure \ref{sphburn}, in addition to the results based on the 100 simulated persistence diagrams. The red (dot dashed )  plot shows the results for the 100 resampled sets from the original persistence diagram, and the yellow (dashed)  plot corresponds to  the 100 resampled sets from the original data. The point where the initial rapid growth of the distance functions ceases is  approximately 22 and 46, respectively,  in Setting I and Setting II for the bottleneck distance, and   approximately 20 and 48   in the  Wasserstein case.
This suggests taking a burn in period of 50 for generating the replicated persistence diagrams for $H_0$.

\subsubsection{$H_1$ persistence diagram}
We now turn to the analysis of the $H_1$ persistence diagram. Again estimating the parameters for the Giibs pseudolikelihood \eqref{eq:pseudo}, taking $K=3$, the estimate of $\delta$ was 0.0047. For this $\delta$, the estimates of $\Theta$ were $\theta _{1} = -0.0331$, $\theta _{2} = 0$, $\theta _{3} = 3.3842$, $\theta _{H} =60.00$, and $\theta _{V} =110.00$.

\begin{figure}[h!]
    \centering
    \subfigure[]
    {
        \includegraphics[scale=.09]{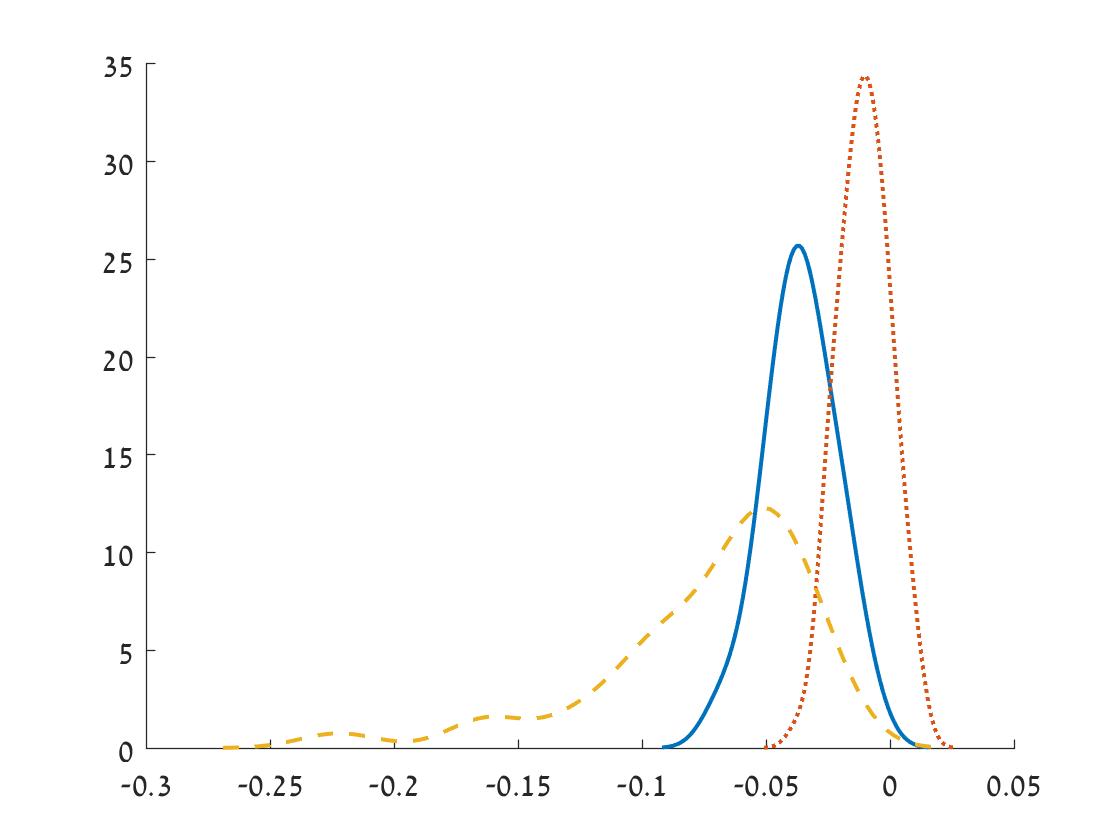}
    }
       \subfigure[]
    {
        \includegraphics[scale=.09]{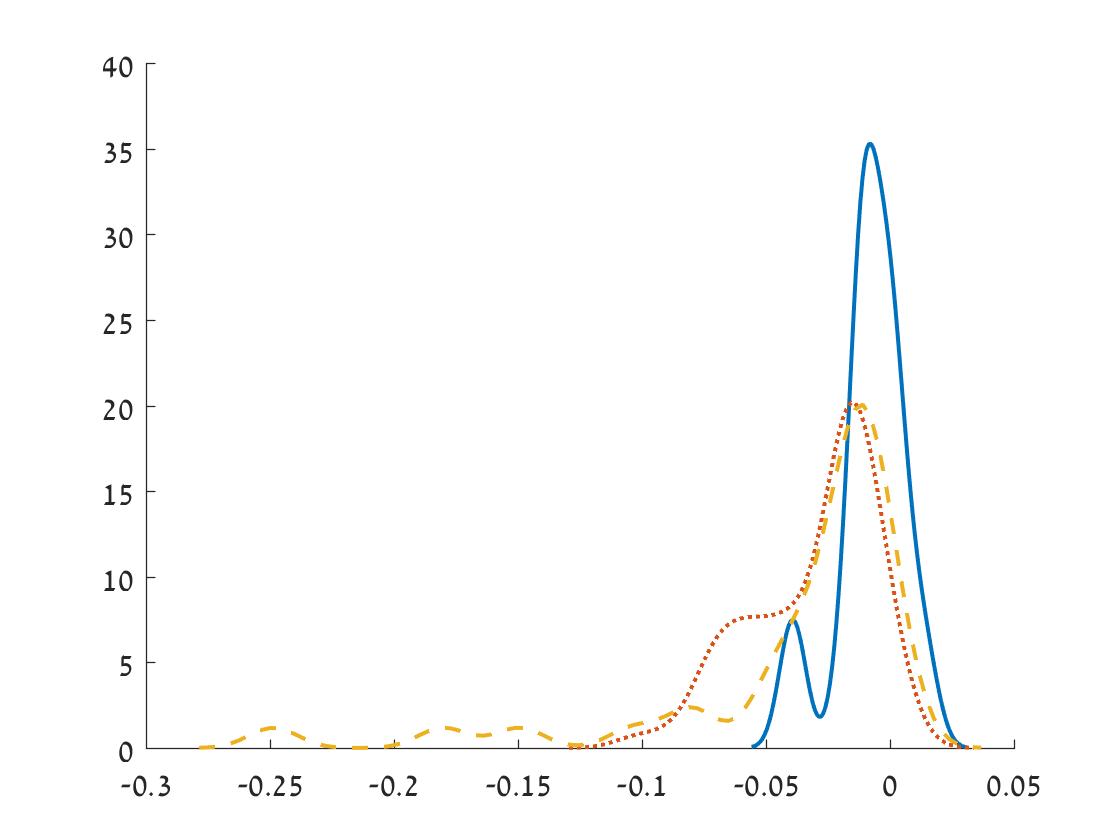}
    }
       \subfigure[]
    {
        \includegraphics[scale=.09]{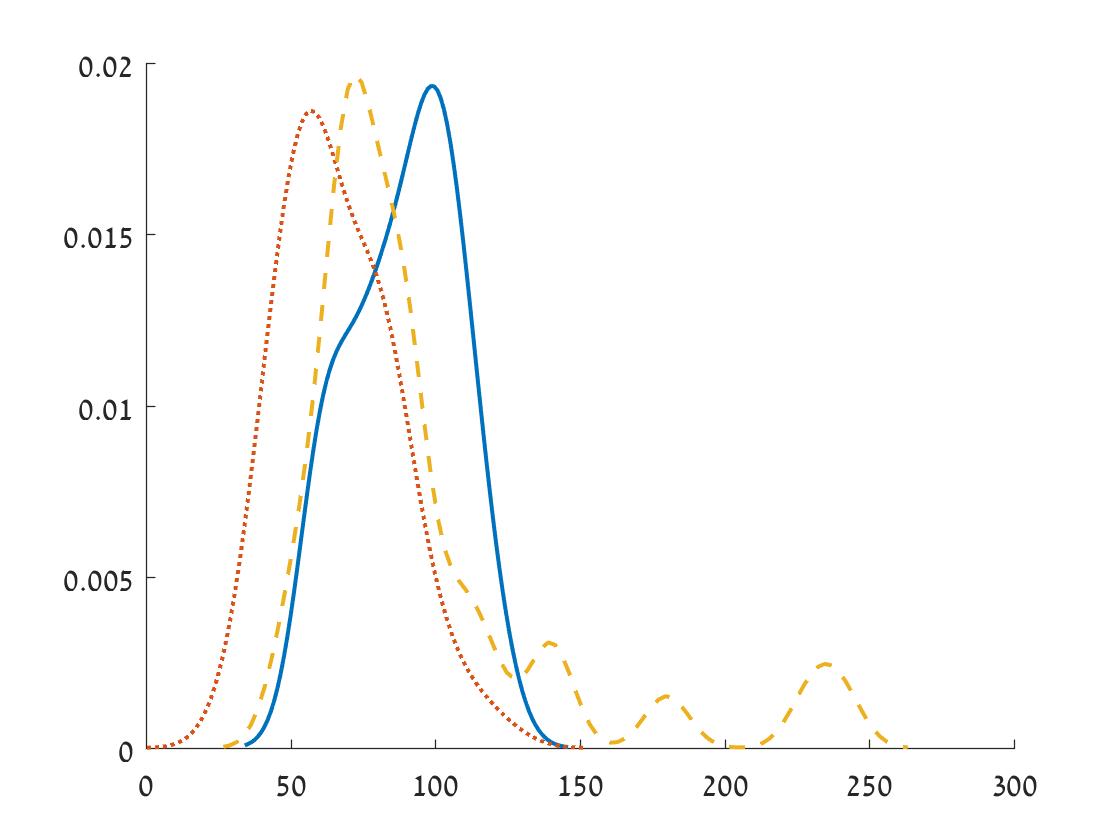}
    }
       \subfigure[]
    {
        \includegraphics[scale=.09]{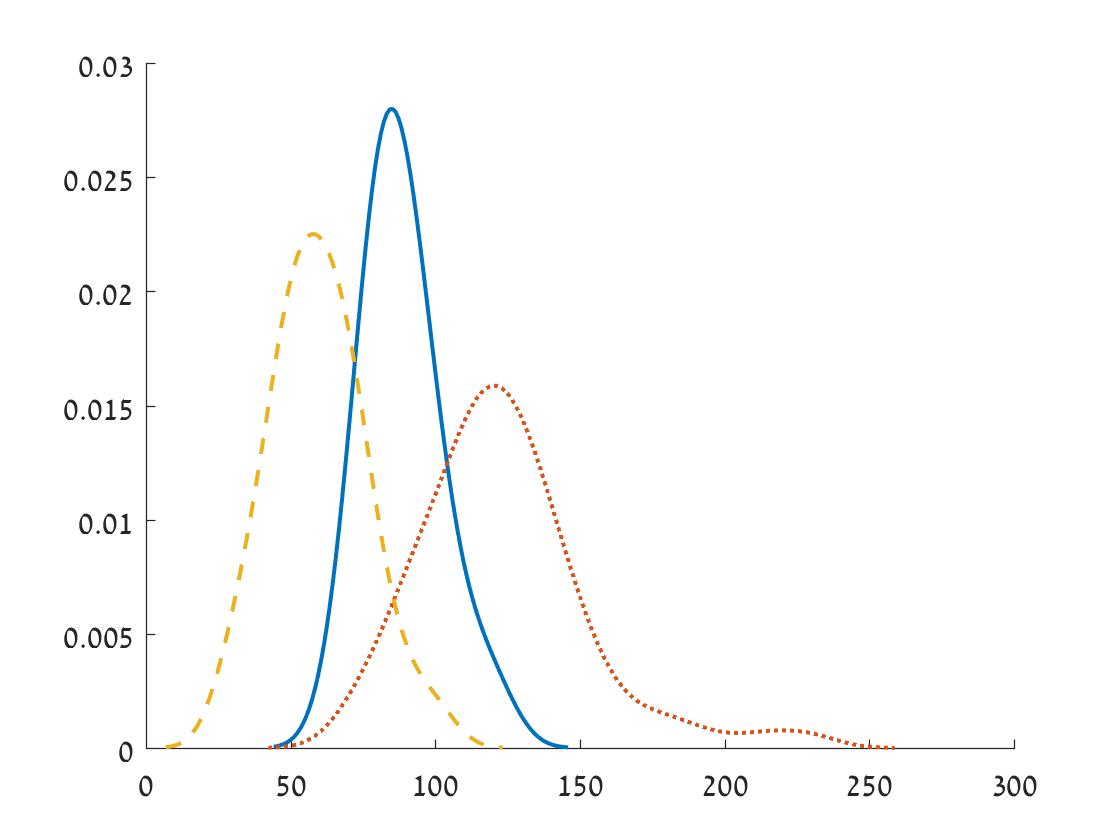}
    }

\caption{\footnotesize
 Smoothed empirical densities for the four parameter estimates of
$H_1$ persistence diagram coming from the simulations of 2-sphere, see text for details. (a) $\theta_1$, (b) $\theta_3$, (c) $\theta_H$, (d) $\theta_V$.}.

\label{fig:resPDestsph1_h1}
\end{figure}

To check the match between the estimated  model  and the $H_1$ persistence diagram, we used the same 100 simulated sets of the  2-sphere used for the $H_0$ diagram, following the same procedure that we adopted then, this time restricting to a model with only
$\theta _{1}$, $\theta _{3}$, $\theta _{H}$, and $\theta _{V}$ non-zero.
The blue plot in Figure  \ref{fig:resPDestsph1_h1} shows the smoothed empirical densities for the parameter estimates generated by these simulations.
%
%
As for the $H_0$ case, the results indicate that the estimation procedure is stable.

\subsubsection{Replicating the  $H_1$ persistence diagram}
\begin{figure}[h!]
    \centering
    \subfigure[]
    {
        \includegraphics[scale=.09]{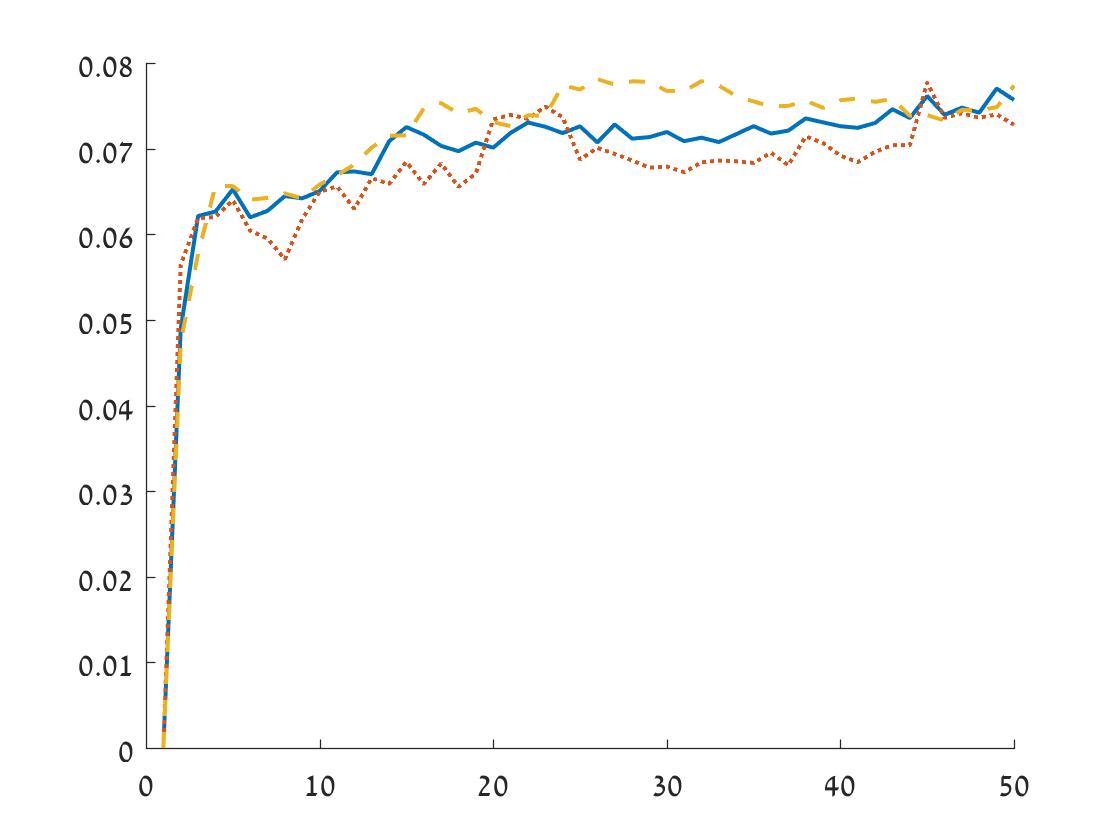}
    }
  \subfigure[]
    {
        \includegraphics[scale=.09]{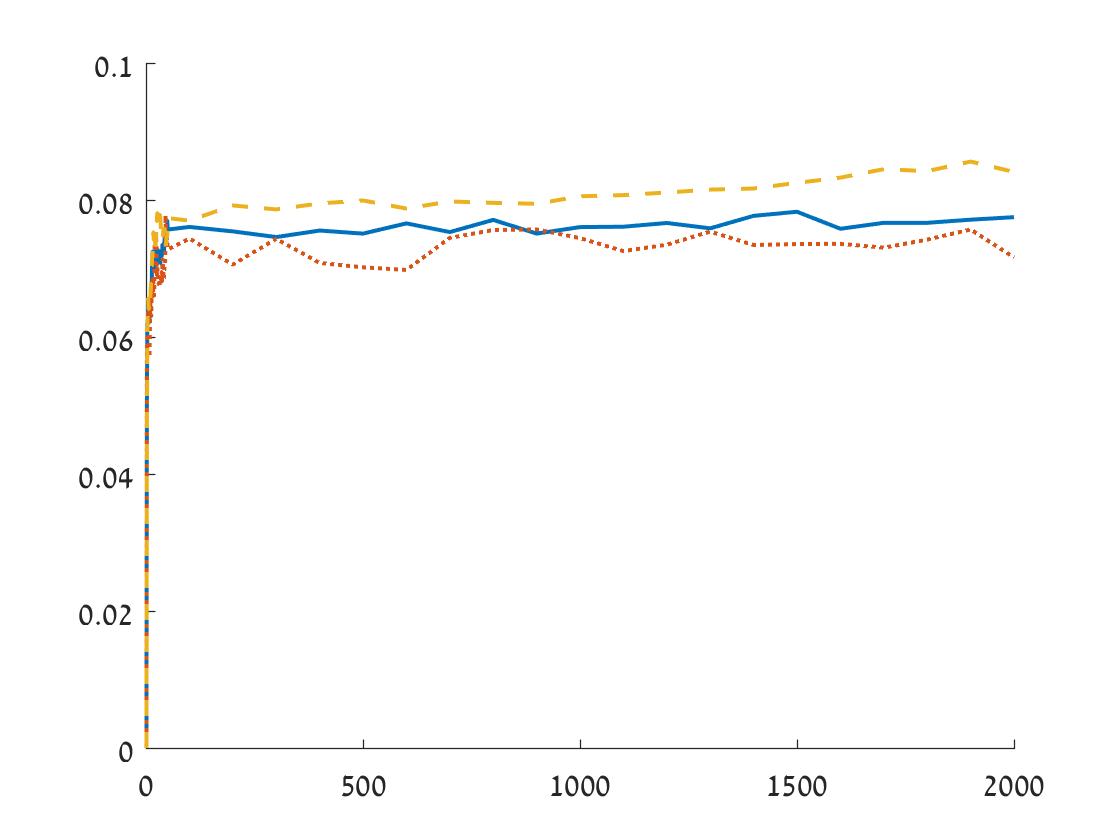}
    }
    \subfigure[]
    {
        \includegraphics[scale=.09]{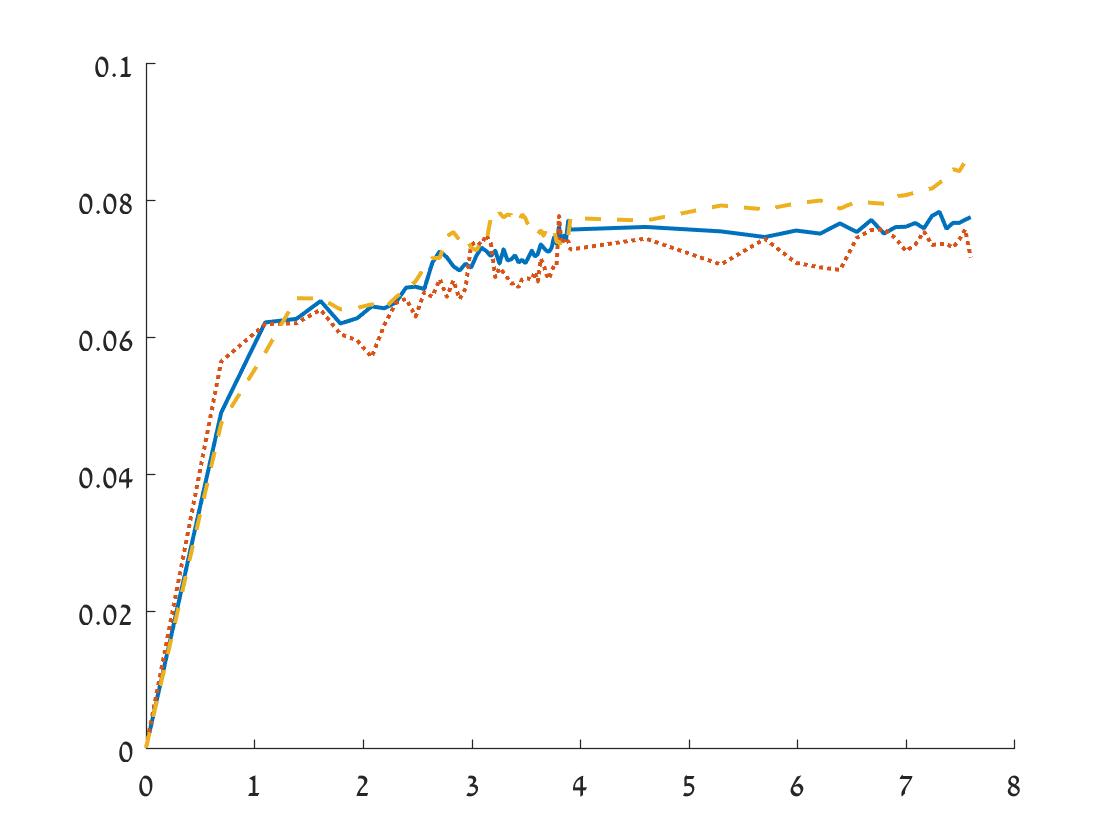}
    }
      \subfigure[]
    {
        \includegraphics[scale=.09]{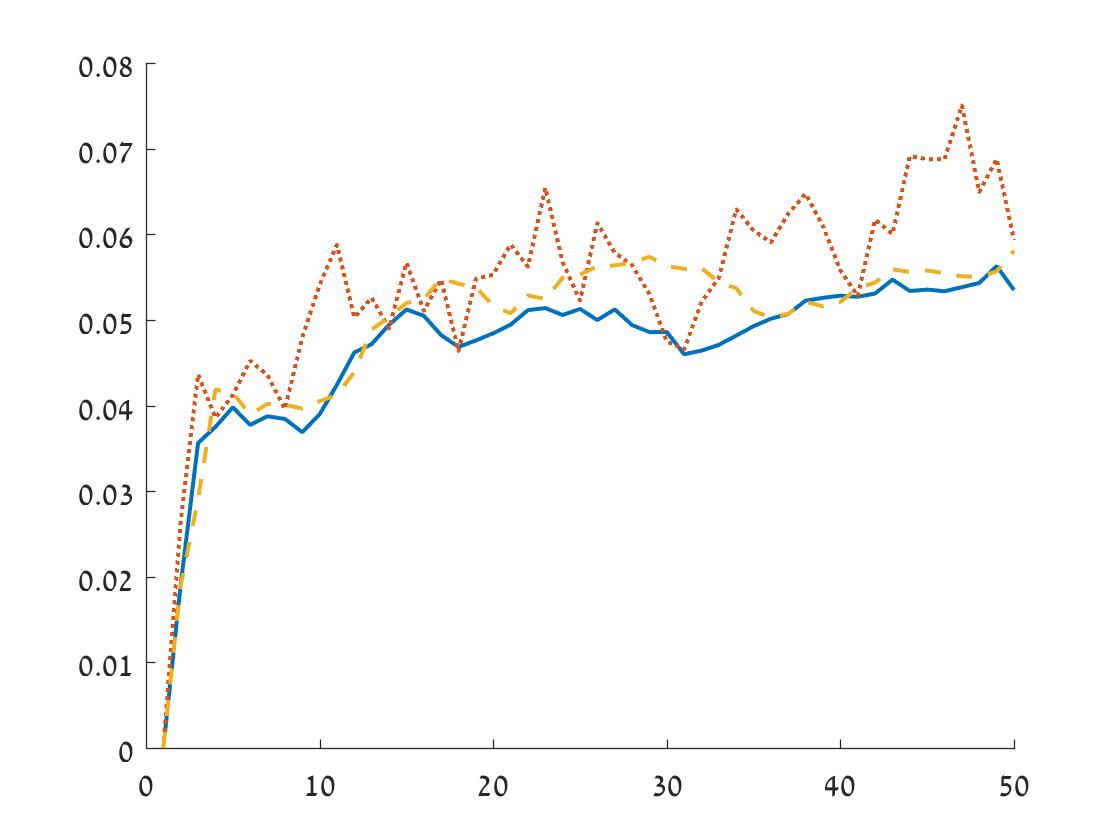}
    }
  \subfigure[]
    {
        \includegraphics[scale=.09]{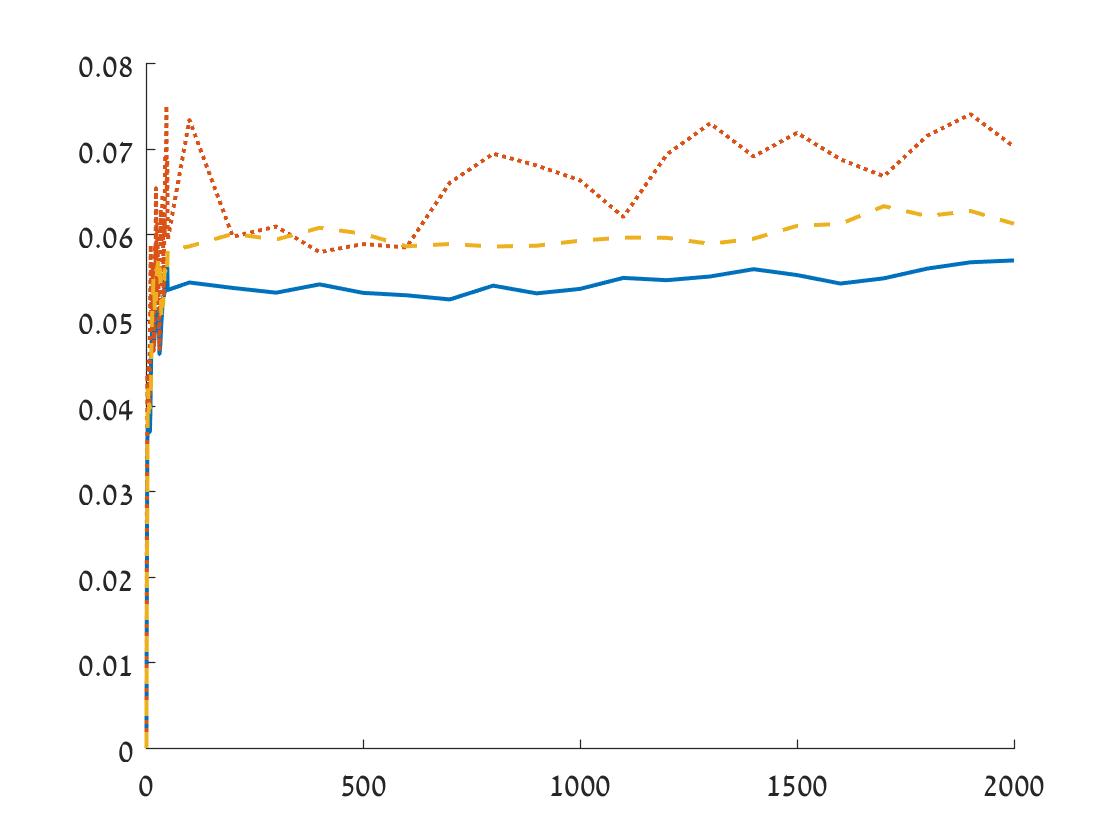}
    }
    \subfigure[]
    {
        \includegraphics[scale=.09]{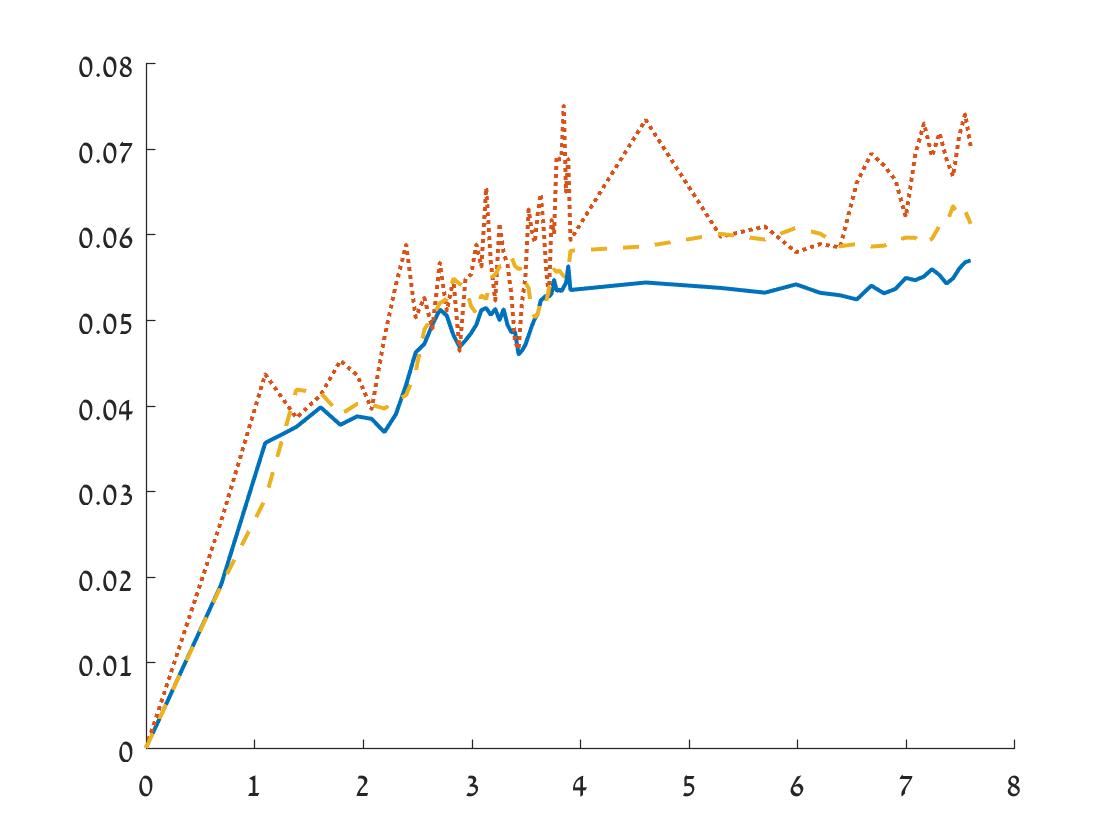}
    }
%
%
\caption{\footnotesize
    Growth of the bottleneck (a) and Wasserstein (d) differences of MCMC simulations from a specific persistence diagram (vertical axis), as a function of the number of steps $n_b$ (horizontal axis, $1\leq n_b\leq 50$)  averaged over 100 independent  persistence diagrams. Panels  (b) and  (e) take $1\leq n_b\leq 2,000$, while (c) and (f) show he same data but on a logarithmic scale.}
\label{fig:burnsphr2}
\end{figure}

As for the analysis of the $H_0$ diagram, we calculated  bottleneck and the Wasserstein distances between the original persistence diagram using those corresponding to 100 MCMC simulated  diagrams of the previous section. The results are shown by the blue plots in Figure  \ref{fig:burnsphr2}.  

The first row in Figure  \ref{fig:burnsphr2} shows the bottleneck distances, while the second row shows the $W_2$ differences. The first column shows the results of the first 50 steps of the MCMC algorithm on a linear scale. The second and third columns go out to 2,000 steps, first on a linear scale and then on a logarithmic scale. The point where the initial rapid growth of the distance functions ceases, is  approximately 44 for the bottleneck distance and 47 in the  Wasserstein case. 

In addition we considered summary statistics of the 100 simulated persistence diagrams as the MCMC progressed, to ensure that  the simulations reliably replicate the statistical properties of the persistence diagrams. Here the best fits were for a burn in of 50, which is consistent with the results of  Figure  \ref{fig:burnsphr2}.
\subsubsection{Resampling $H_1$}
As for the $H_0$ case, we again examine the performance of resampling  from the original persistence diagram (Setting I) and    from the original data (Setting II), repeating each procedure  100 times.
The results are summarised in  Figure  \ref{fig:resPDestsph1_h1}. The red (dot dashed) plots are the smoothed empirical densities for the parameter estimates in Setting I, while  the yellow (dashed) plot correspond to Setting II.
%


In order to assess the fit of the simulated data to the original, we computed, as previously, the bottleneck and the Wasserstein distances between the MCMC simulations and the data itself. The results are presented in Figure \ref{fig:burnsphr2}, in addition to the results based on the 100 simulated persistence diagrams. The red (dot dashed )  plot shows the results for the 100 resampled sets from the original persistence diagram, and the yellow (dashed)  plot shows the same thing, but for the 100 resampled sets from the original data. The point where the initial rapid growth of the distance functions ceases, is  approximately 22 and 46 in Setting I and Setting II, respectively, for the bottleneck distance, and   approximately 20 and 48   in the  Wasserstein case.
This suggests taking a burn in period of 50 for generating the replicated persistence diagrams for $H_1$.

\subsubsection{Statistical inference}
We are now finally in  a position to carry out a simulation study to test how well we can identify the homology of 2-sphere, using the methodology described earlier. To do so, we
generated  1,000 persistence diagrams from the fitted model, via MCMC, with a burn in period of 50 iterations and with  $(n_b,n_r,n_R)$ given by (500,10,100), (500,20,50), (500,40,25),  or (500,100,10).
Using these four sets of simulations, we computed the maximum statistics $T_1$, its confidence interval and its $p$-value, for both the $H_0$ and $H_1$ persistence diagrams. Table \ref{table:spherestat} summarizes the results.

\begin{table}[h!]
\begin{center}
\fontsize{8.5}{0.9}\selectfont
\begin{tabular}{llccccc}

homology&statistic&real PD &$(n_b, n_r, n_R)$ & CI & $p$-value& significance\\
\\
\\
\\
\\
$H_0$&$T_1$& 0.4769 &(500,10,100)&[0, 0.4769]&0.0990    &	no\\
\\
     & & &(500,20,50)&[0, 0.4769]&0.0520     &	no\\
\\
     & & &(500,40,25)&[0, 0.3273]&0.0320     &	yes\\
\\
     & & &(500,100,10)&[0, 0.2616]&0.0100  &yes\\
\\
$H_1$&$T_1$& 0.1673 &(500,10,100)&[0, 0.2140]&0.4060    &	no\\
\\
     & & &(500,20,50)&[0, 0.2069]&0.3780     &	no\\
\\
     & & &(500,40,25)&[0, 0.2065]&0.3550     &	no\\
\\
     & & &(500,100,10)&[0, 0.1995]&0.3270  &no\\
\\
\\
\\
\\

\end{tabular}
\end{center}
\caption{{ {\footnotesize  Maximum statistic $T_1$ for the real $H_0$ and $H_1$ persistence diagram and the simulated $H_0$ and $H_1$ persistence diagrams of the 2-sphere. The CI is a one-sided confidence interval at a $5\%$ confidence level. The $p$-value is also  one-sided. Both the CI and the $p$-value are based on 1,000 simulated persistence diagrams.}}}

\label{table:spherestat}
\end{table}

\normalsize
The results for the $H_0$ persistence diagram show that $T_1$, in two first scenarios, was statistically insignificant, and in the two other scenarios  was significant. In other words, the evidence is split between one connected component (represented by   the `point  at infinity' not included in the analysis) and two components. The fact that the correct result occurs in the cases of a larger number of shorter MCMC runs is consistent with earlier findings in \cite{adleragamprat}.

%

As for the  $H_1$ topology, all four scenarios showed that $T_1$ was insignificant for  all
MCMC parameter, implying, correctly, a trivial $H_1$ homology.



\subsection{2-torus}
We now turn to our second example, that of the two-dimensional torus.
Since the analysis will be similar in approach to that for the two-dimensional sphere, we will give fewer details, concentrating primarily on the more important differences in the results.

\subsubsection{The data and fitting the model}
\begin{figure}[h!]
    \centering
    \subfigure[]
    {
        \includegraphics[width=1.3in, height=1.3in]{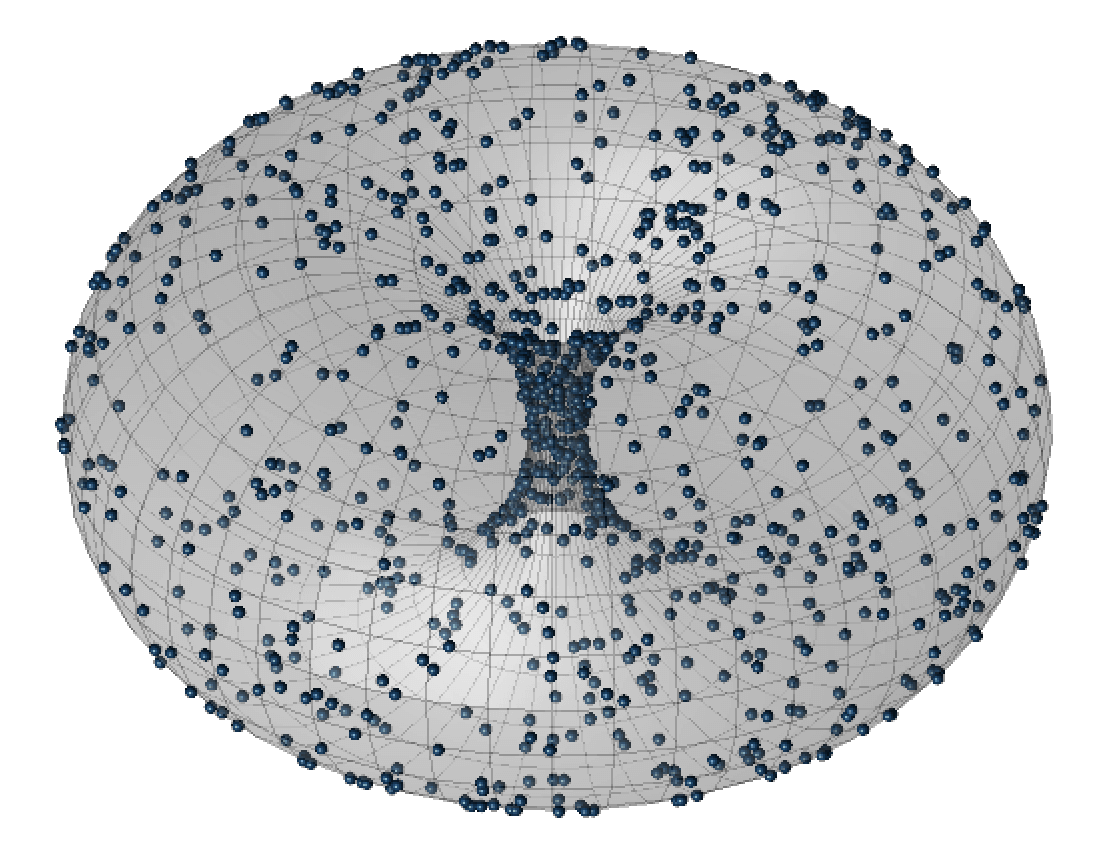}  \ \ \ \
    }
       \subfigure[]
    {
        \includegraphics[width=1.25in, height=1.25in]{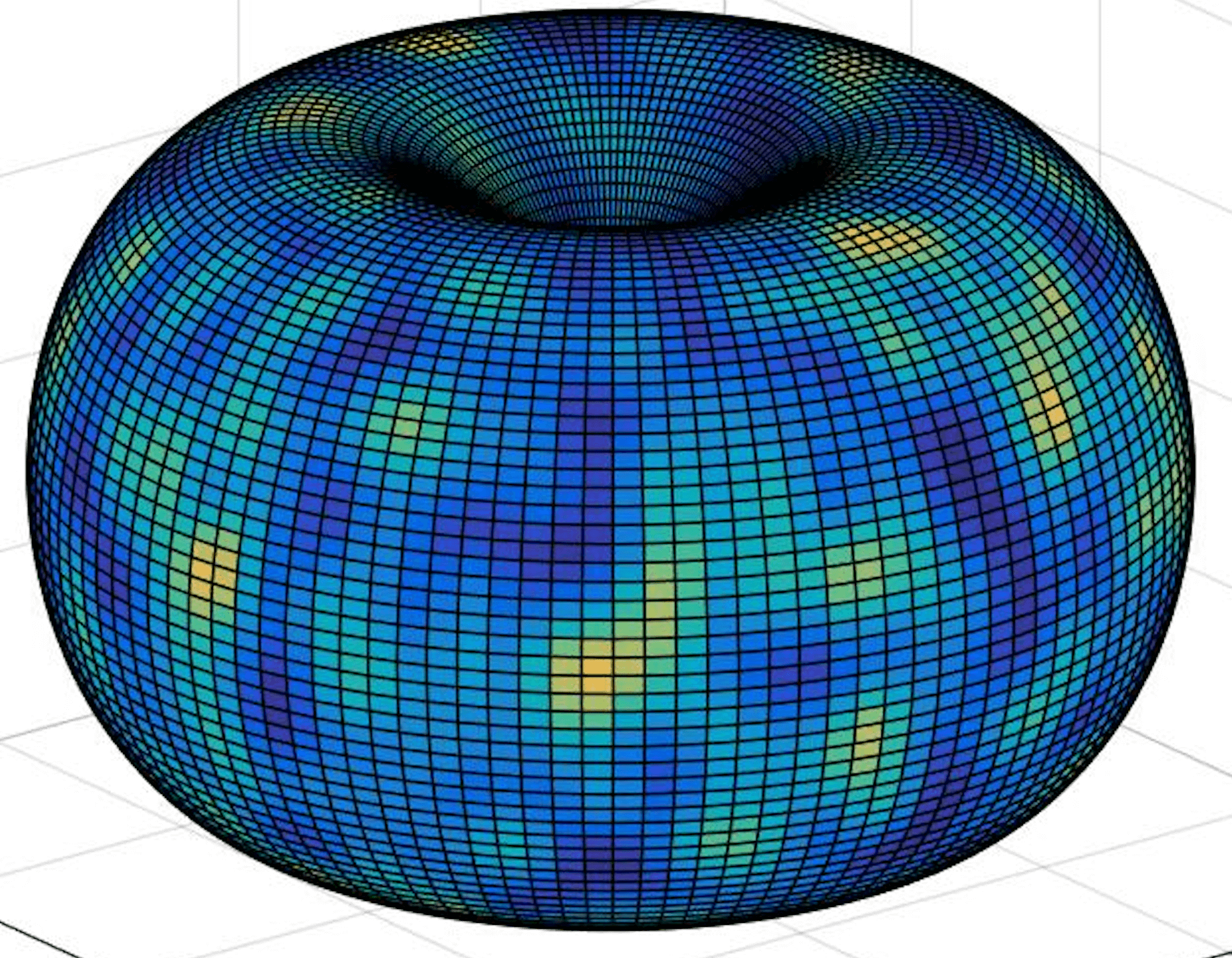} \ \ \ \
    }
       \subfigure[]
    {
        \includegraphics[width=1.3in, height=1.3in]{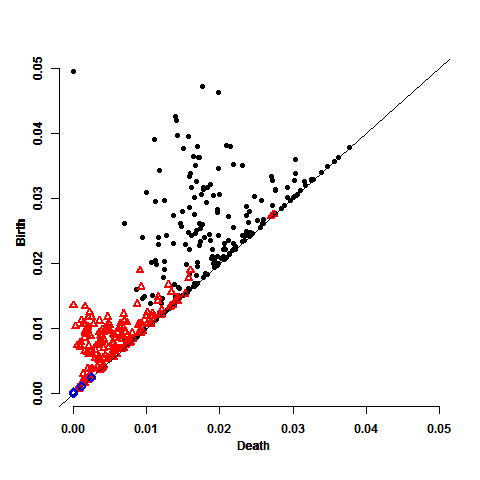}
    }
\caption{\footnotesize
  (a) Points sampled from a torus. View from above the torus. (b) The corresponding kernel density estimator, shown, for visual clarity, at only a few quantized levels, and from a different angle to (a). (c)  The corresponding persistence diagram for the upper level sets of the kernel density estimate on the full torus.  Black circles are  $H_0$ persistence points,  red triangles are  $H_1$ points, blue diamonds are $H_2$ persistence points. Birth times are on the vertical axis.}
\label{fig:torus}
\end{figure}

This example includes a sample of $n=1,000$ points from the 2-torus $T^2=S^1\times S^1$ in $R^3$, chosen uniformly with respect to the natural Riemannian metric induced on it as a subset on $\mathbb R^3$. This leads to the high density of points in the `interior' of the torus, obvious from  Figure  \ref{fig:torus}. (For more details on sampling from tori and other manifolds, see \cite{Persi}.)
More specifically, the  torus was taken to be the rotation about the `$z$ axis' in $\mathbb R^3$ of a circle  of radius $1.8$ with center in the `$(x,y)$ plane'  at distance 2 from the origin.

Panel (a) in  Figure  \ref{fig:torus} shows the sample superimposed on the torus, and Panel (b) shows the corresponding
 kernel density estimator based on a bandwidth of $\eta=0.2$.
The corresponding persistence diagram of the upper level set filtration of $\hat f_N$ is Panel (c).  This diagram contains $N_0=216$ points of the zeroth homology $H_0$, represented by  the black circles, $N_1=160$ points of the first homology $H_1$, represented by  the red triangles, and $N_2=216$ points of the second homology $H_2$, represented by the blue diamonds. Since we know that the upper level sets of $\hat f_N$ are characterized by having a single connected component, two holes, and a single void, we expect to have one black circle and two red triangles somewhat isolated from the other points in the diagram, and one blue diamond. In fact, we can see the one isolated black circle point, but it is not clear which are the two main red triangles.

Adopting the approach described above separately for the $H_0$ and $H_1$ persistence diagrams, we estimated the parameters for a Gibbs distribution for the model with pseudolikelihood \eqref{eq:pseudo}, taking $K=3$.
For the $H_0$  diagram, working without  the point at infinity, the estimate of $\delta$ was 0.0010, and the estimates of $\Theta$ were $\theta _{1} = -0.0034$, $\theta _{2} = -0.0026$,
$\theta _{3} = -0.0032$, $\theta _{H} =1.08E+04$, and $\theta _{V} =4.19E+03$.

For the $H_1$ persistence diagram, the estimate of $\delta$ was 0.0007, and the estimates of $\Theta$ were $\theta _{1} = -0.0044$, $\theta _{2} = -0.0059$,
$\theta _{3} = -0.0036$, $\theta _{H} =1.29E+05$, and $\theta _{V} =1.50E+04$.

\subsubsection{Replicating the persistence diagram}
\begin{figure}[h!]
\bc
\includegraphics[width=0.9in, height=0.9in]{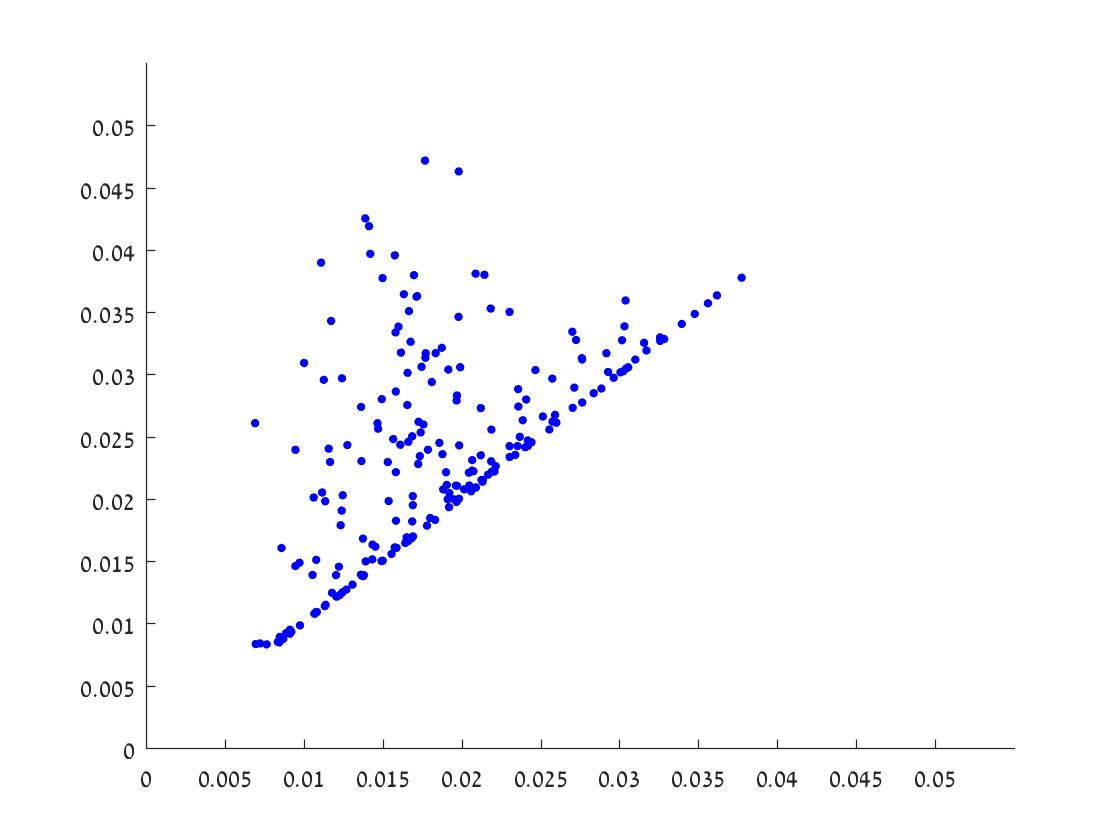}
\includegraphics[width=0.9in, height=0.9in]{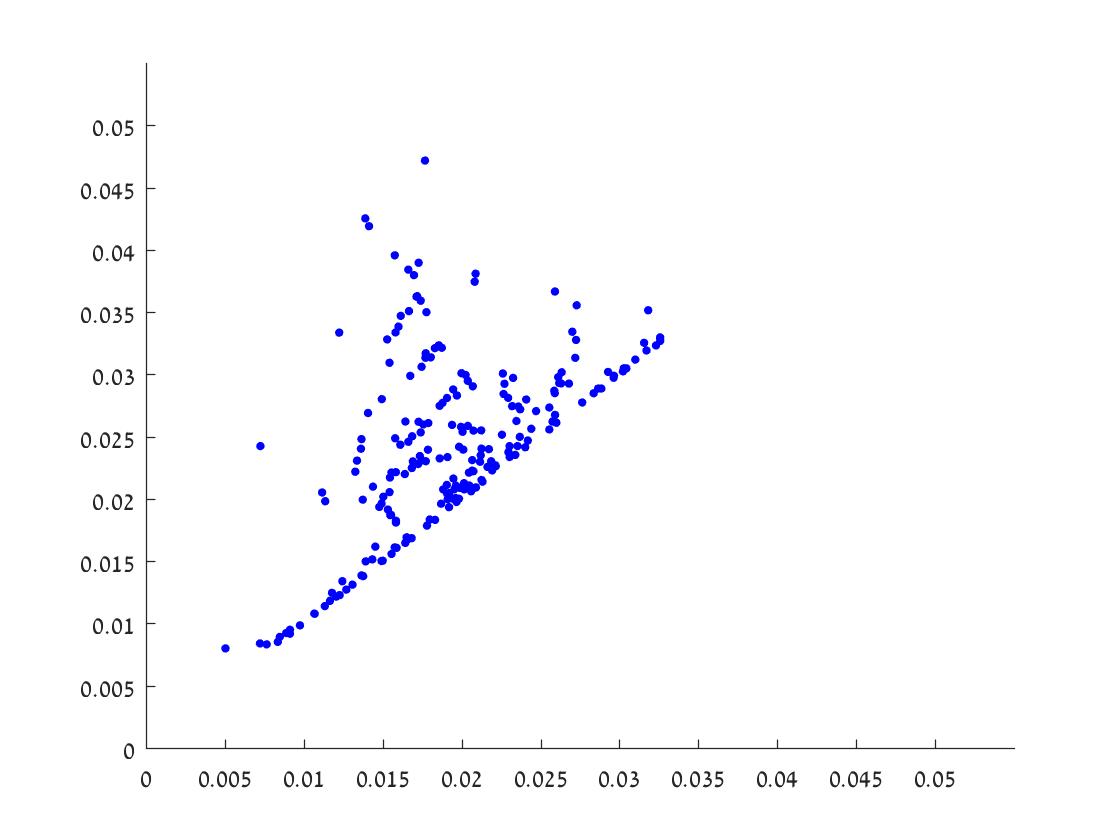}
\includegraphics[width=0.9in, height=0.9in]{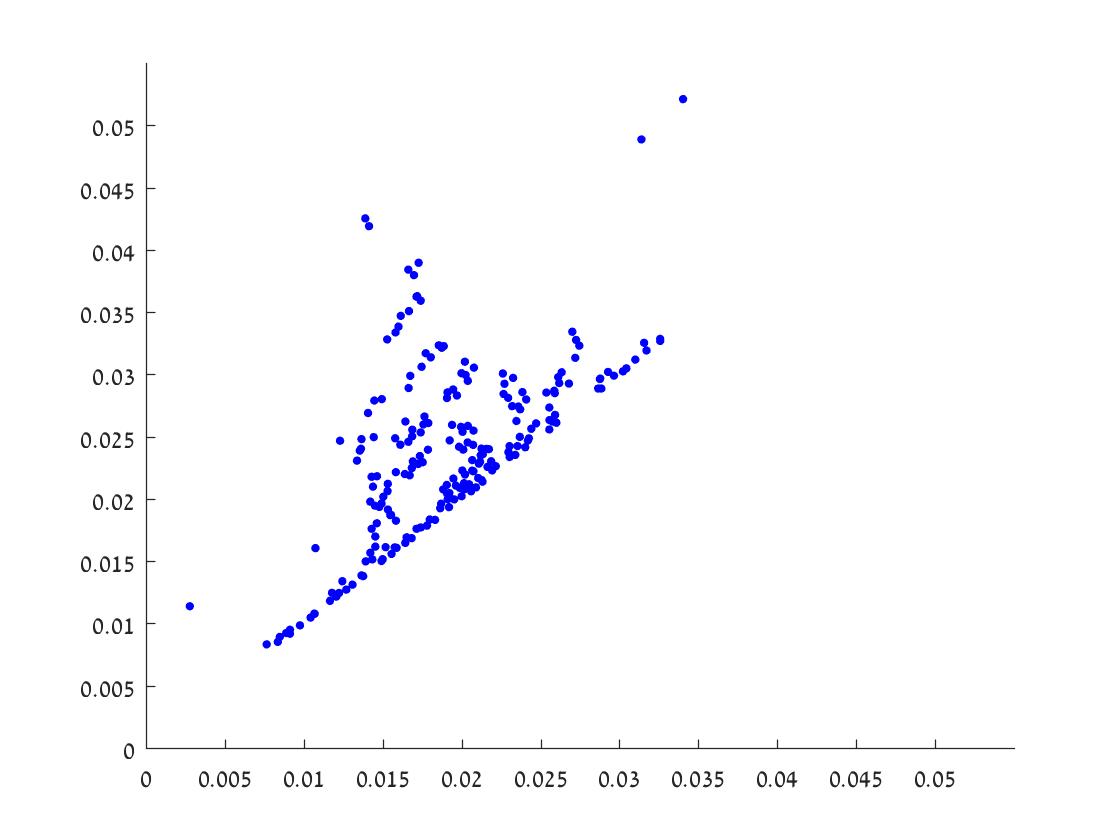}
\includegraphics[width=0.9in, height=0.9in]{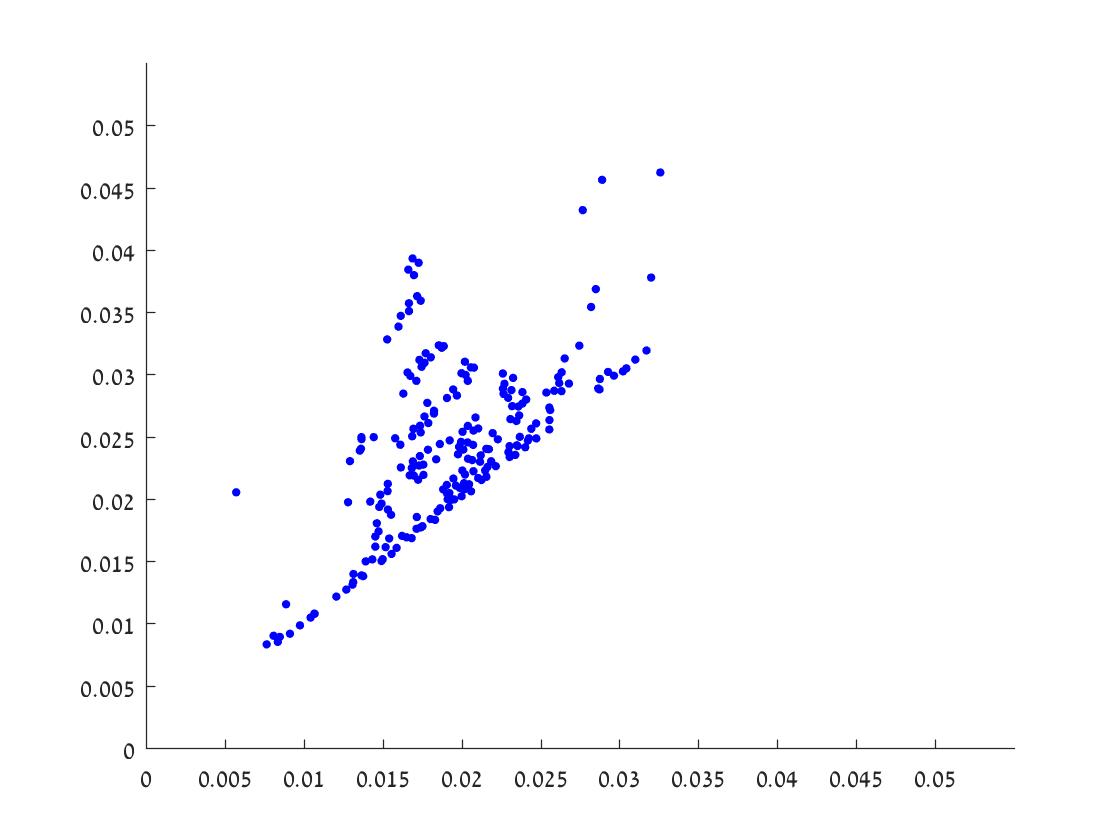}
\includegraphics[width=0.9in, height=0.9in]{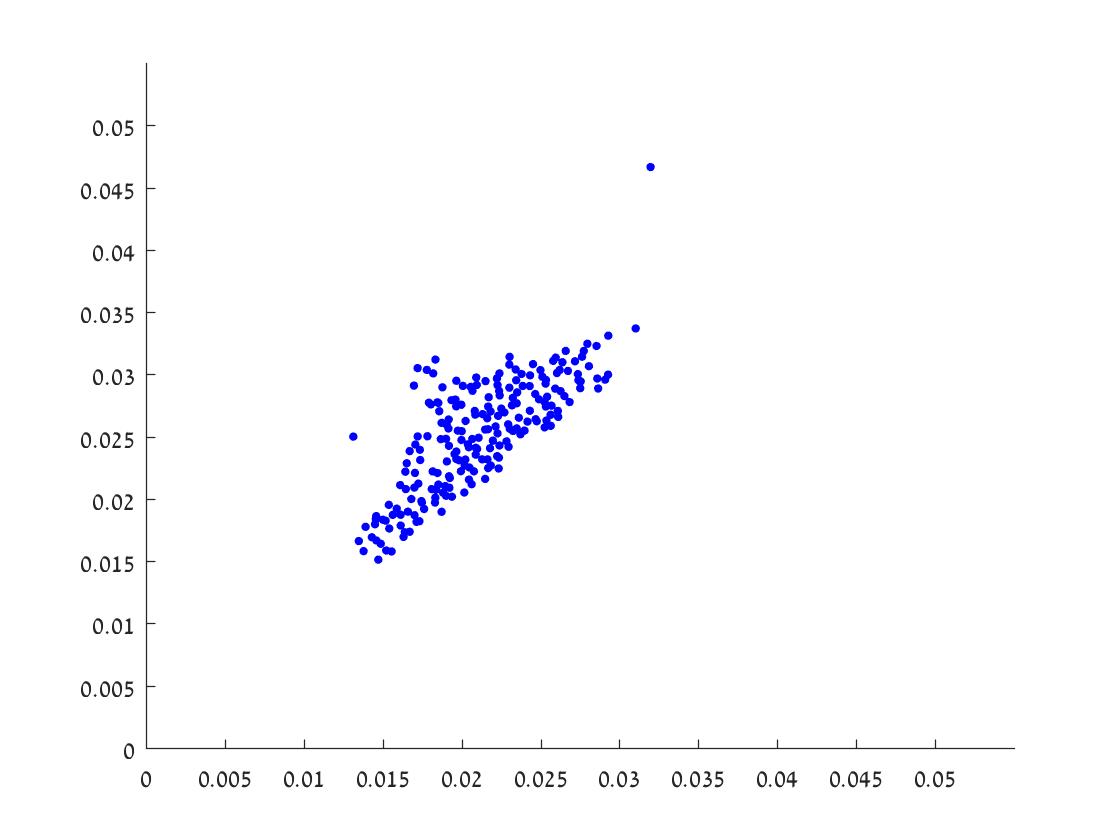}
\includegraphics[width=0.9in, height=0.9in]{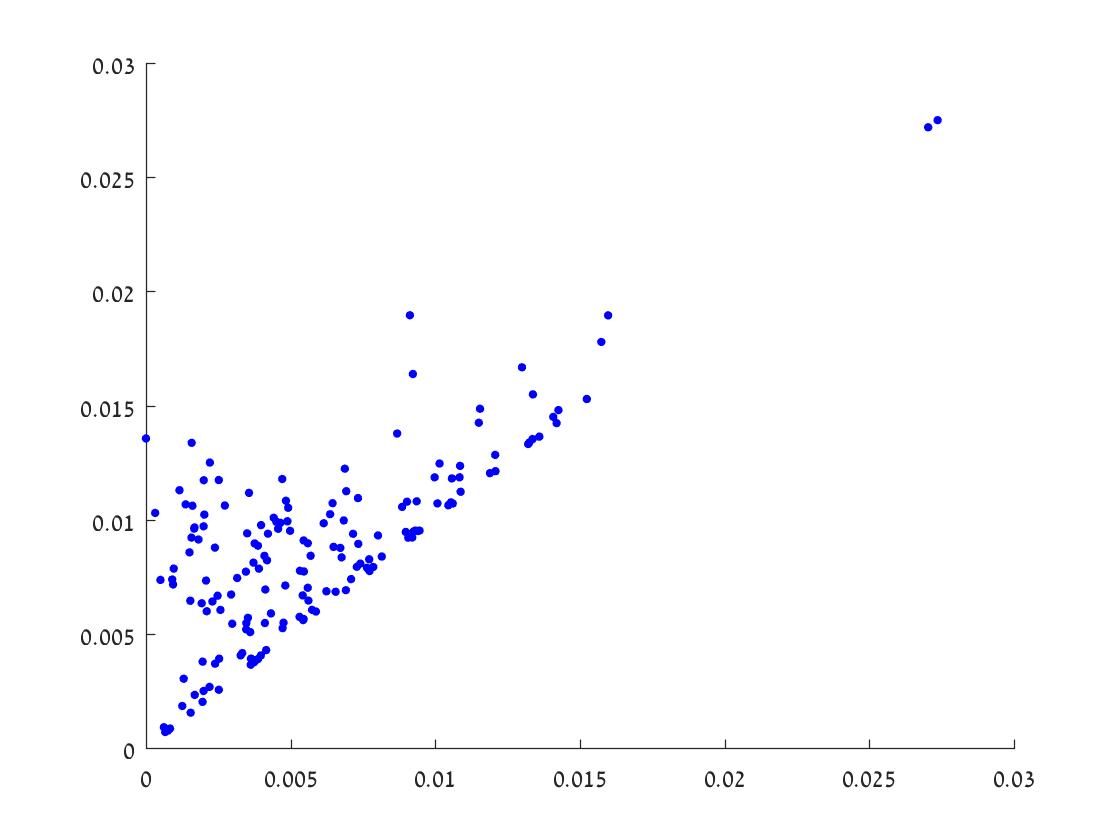}
\includegraphics[width=0.9in, height=0.9in]{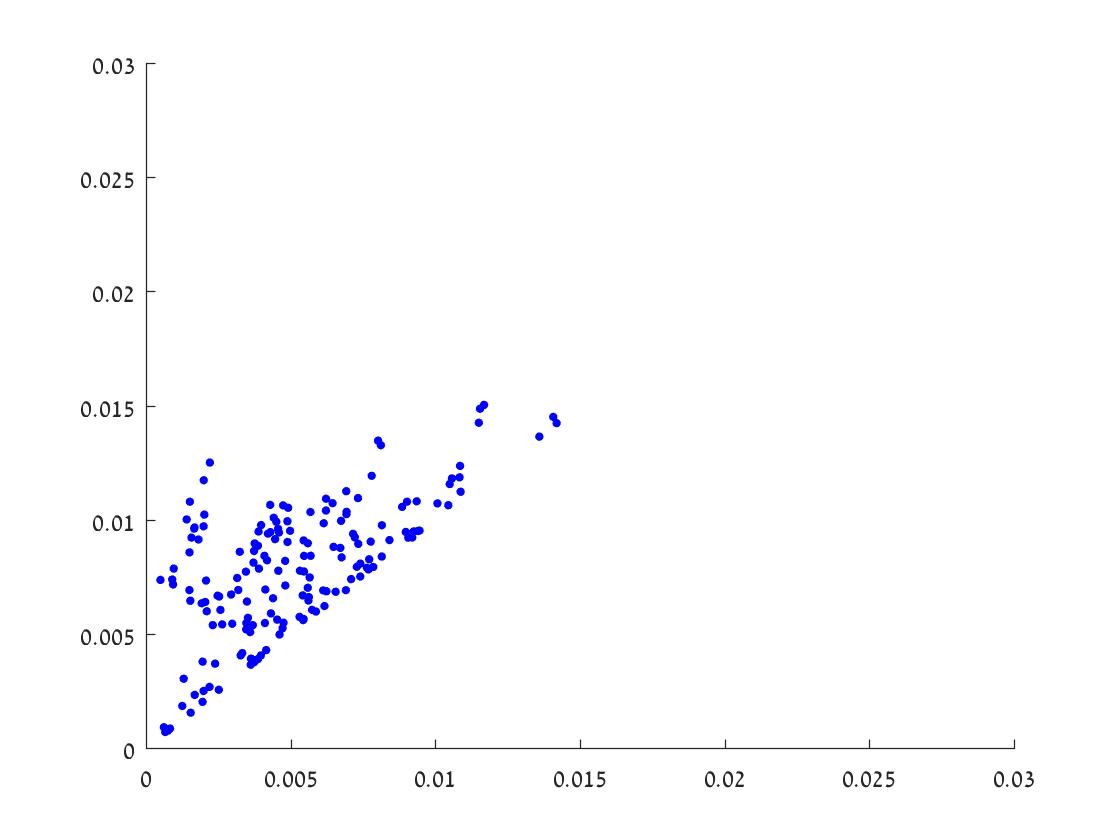}
\includegraphics[width=0.9in, height=0.9in]{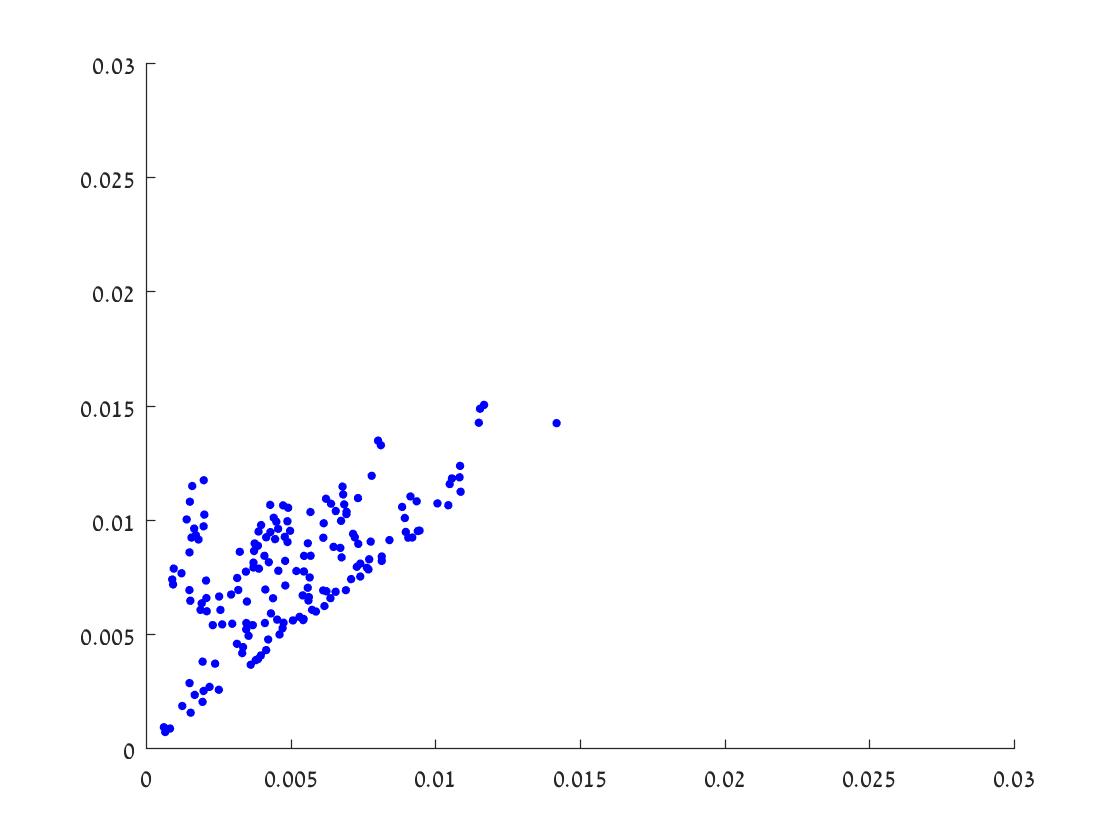}
\includegraphics[width=0.9in, height=0.9in]{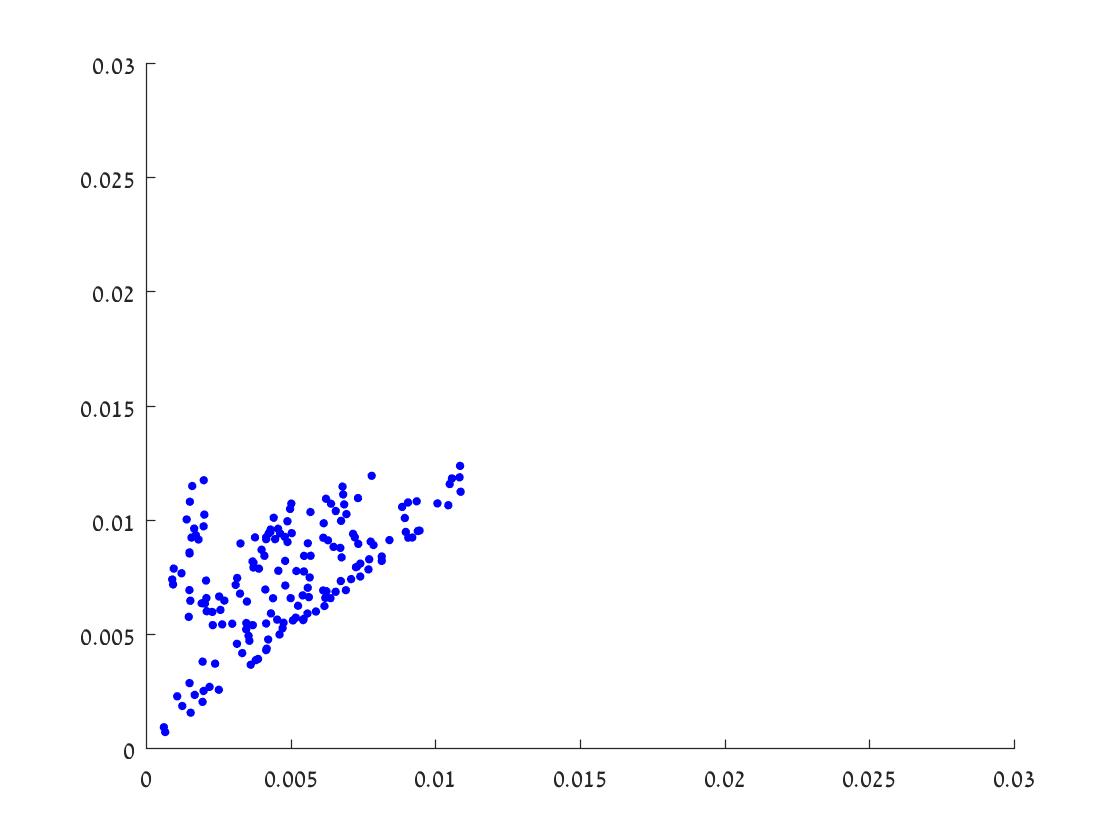}
\includegraphics[width=0.9in, height=0.9in]{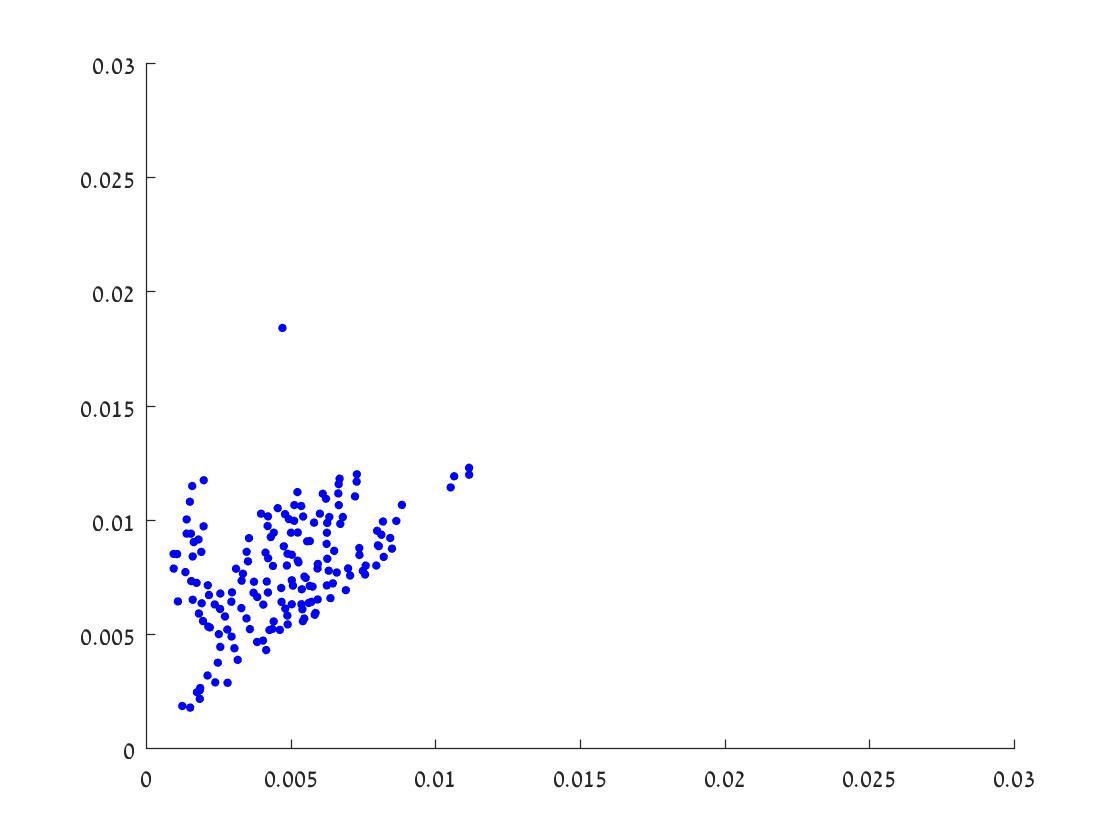}
\ec
\caption{\footnotesize
The first row shows the persistence diagrams of $H_0$, and the second row shows the persistence diagrams of $H_1$.
At each row, the left plot is the original persistence diagram, and the four other plots are simulated persistence diagrams based on an MCMC simulation with burn in of 10, 25, 50 and 1000. }
\label{fig:torusmcmc}
\end{figure}

The determination of the burn in period in this example, for both $H_0$ and $H_1$, was only heuristic. Figure  \ref{fig:torusmcmc} presents the original persistence diagrams of $H_0$ and $H_1$ and their MCMC with burn in periods of 10, 25, 50 and 1000. The best fits for both $H_0$ and $H_1$ occur for burn in periods in the range $[10,50]$.

\subsubsection{Statistical inference}
We generated  1,000 replicated persistence diagrams from the fitted model with a burn in period of 10 iterations and with  $(n_b,n_r,n_R)$ given by (500,10,100), (500,20,50), (500,40,25), or (500,100,10).
Using these four sets of simulations, we computed the maximum statistics $T_j$, $j=1,...,4$, their confidence intervals and their $p$-values. Table \ref{table:torus} summarizes the results.

\begin{table}[h!]
\begin{center}
\fontsize{8.5}{0.9}\selectfont
\begin{tabular}{llccccc}

homology&statistics&real PD &$(n_b, n_r, n_R)$ & CI & $p$-value& significance\\
\\
\\
\\
$H_0$&$T_1$& 0.0295 &(500,10,100)&[0, 0.0371]&0.3360     &no\\
\\
     & &       &(500,20,50)&[0, 0.0362]&0.2770     &	no\\
\\
     & &       &(500,40,25)&[0, 0.0359]&0.2350     &	no\\
\\
     & &       &(500,100,10)&[0, 0.0322]]&0.2720     &	no\\
\\
\\
\\
\\
$H_1$&$T_1$& 0.0136 &(500,10,100)&[0, 0.0123]&0.0340     &	yes\\
\\
      &&       &(500,20,50)&[0, 0.0118]&0.0320     &	yes\\
\\
      &&       &(500,40,25)&[0, 0.0107]&0.0220     &	yes\\
\\
      &&       &(500,100,10)&[0, 0.0134]]&0.0490     &	yes\\
\\
\\
\\
\\
$H_1$&$T_2$& 0.0118 &(500,10,100)&[0, 0.0103]&0.0030     &	yes\\
\\
      &&       &(500,20,50)&[0, 0.0102]&0.0060     &	yes\\
\\
      &&       &(500,40,25)&[0, 0.0102]&0     &	yes\\
\\
      &&       &(500,100,10)&[0, 0.0101]]&0.0020     &	yes\\
\\
\\
\\
\\
\\
$H_1$&$T_3$& 0.0103 &(500,10,100)&[0, 0.0100]&0.0360     &	yes\\
\\
      &&       &(500,20,50)&[0, 0.0098]&0.0060     &	yes\\
\\
      &&       &(500,40,25)&[0, 0.0099]&0.0060     &	yes\\
\\
      &&       &(500,100,10)&[0, 0.0099]]&0.0180     &	yes\\
\\
\\
\\
\\
\\
\label{table:torus}
\end{tabular}
\end{center}
\caption{{ {\footnotesize  Maximum statistics $T_1$, $T_2$, $T_3$ for the real $H_0$ and $H_1$ persistence diagrams and the simulated $H_0$ and $H_1$ persistence diagrams of the 2-torus. The CI is a one-sided confidence interval at a $5\%$ confidence level. The $p$-value is also a one-sided. Both the CI and the $p$-value are based on 1000 simulated persistence diagrams.}}}

\label{table:torus}
\end{table}
The results for the $H_0$ diagram, for all scenarios, showed that $T_1$ was insignificant (the lowest $p$-value
reached in any of the six cases was 0.235). Thus, adding the `point at infinity' back into the diagram, we have evidence for       exactly one connected component, as we hoped to find.

For the $H_1$ diagram, the results for all scenarios showed that $T_1$ and $T_2$ were significant (the highest $p$-value
reached in any of the 8 cases was 0.049). That is, two significant holes, as we hoped to find.

Unfortunately, however,  $T_3,...,T_8$ were also statistically significant,  leading to a significantly over-estimation of the complexity of the $H_1$ homology. This seems to be due to the sparsity of points in the sample, which is clear from the first two panels in Figure \ref{fig:torus}. Our conclusion here, therefore, is a need for either a larger sample size or, perhaps, a larger bandwidth for the kernel density estimator.

\subsection{Three circles}

Our final example is that of three concentric circles in $\mathbb R^2$. We describe the main results, skimping on detail.

\subsubsection{The data and fitting the model}
\begin{figure}[h!]
\bc
    \subfigure[]
    {
        \includegraphics[width=1.3in, height=1.3in]{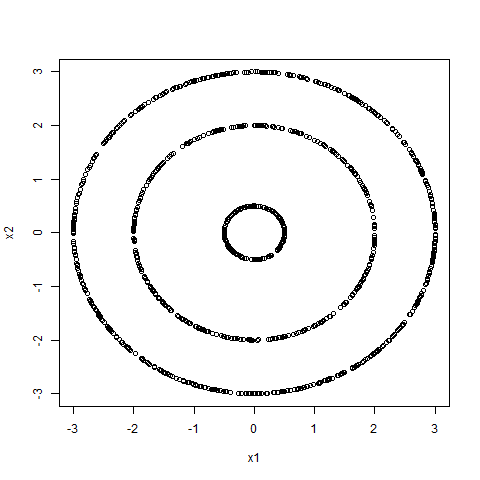}
    }
       \subfigure[]
    {
        \includegraphics[width=1.3in, height=1.3in]{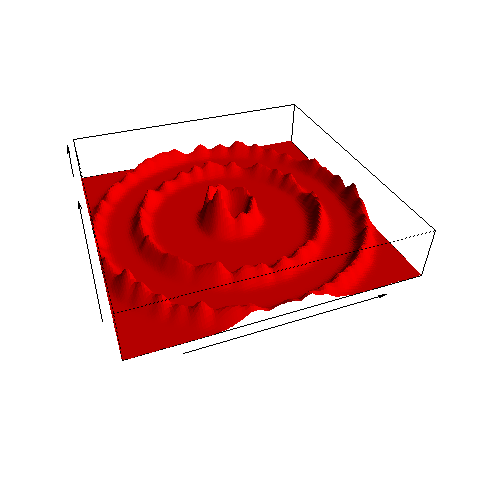}
    }
       \subfigure[]
    {
        \includegraphics[width=1.3in, height=1.3in]{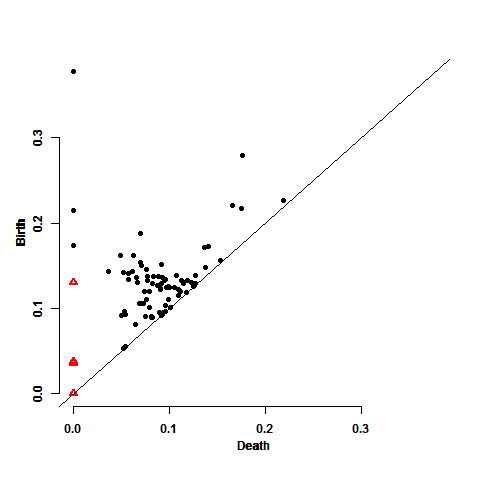}
    }

\ec
\caption{\footnotesize
 (a) A random sample from three circles, 600 points from the larger circle, 400 points from the middle circle, and 200 from the smaller one,  with a kernel density estimate (b) and the persistence diagram (c) for its upper level sets. Black circles are  $H_0$ persistence points,  red triangles are  $H_1$ points. Birth times are on the vertical axis.}

\label{fig:threecirc}
\end{figure}

For this example we start with a random sample of $n=1,200$ points from three circles, of diameters 6, 4 and 1, as shown in the  Panel (a) of Figure \ref{fig:threecirc}.
In total,  600 points were chosen  from the largest circle, 400 from the middle circle, and 200 from the smallest one.  Panel (b)  shows  the corresponding kernel density estimate,  for which we took the bandwidth  $\eta =0.1$. Panel (c) displays the corresponding persistence diagram of the upper level set filtration of $\hat f_N$, containing $N_0=80$ points of the zeroth homology $H_0$, represented by  the black circles,  with the red triangles corresponding to the first homology $H_1$.  Since we know that the upper level sets of $\hat f_N$ are characterized by having three main components, each of which contains a single 1-cycle (hole) we expect to see three black circles and three red triangles somewhat isolated from the other points in the diagram, which  is in fact the case.

Note firstly that while there are quite a few (black, circular) points corresponding to the $H_0$ homology, there are only four (red, triangular)  for $H_1$. Since the methodology described in the previous sections requires the estimation of a sophisticated model with a number of parameters, it follows that it is not appropriate for modelling the $H_1$ part of the diagram.  However, there are more than  enough $H_0$ points in Figure  \ref{fig:threecirc} to fit a spatial model to them.

Adopting the approach described above, and working only with the $H_0$ persistence diagram without including the  `point at infinity', we estimated the parameters for a Gibbs distribution for the model with pseudolikelihood \eqref{eq:pseudo}, taking $K=3$. The estimate of $\delta$ was 0.0012. For this $\delta$, the estimates of $\Theta$ were $\theta _{1} = -0.0105$, $\theta _{2} = 0$,
$\theta _{3} = 0$, $\theta _{H} =394.6$, and $\theta _{V} =147.7 $.
\subsubsection{Statistical inference}
We computed, as previously, the bottleneck and the Wasserstein distances between the MCMC simulations and the data itself, based on 100 simulated sets that behave the same as our original data. The point where the initial rapid growth of the distance functions ceases was  approximately 10 for the bottleneck distance and was  approximately 15 in the  Wasserstein case.
Based on these results, we
generated  1,000 persistence diagrams from the fitted model with a burn in period of 10 iterations and with  $(n_b,n_r,n_R)$ given by (500,20,50), (500,40,25),  or (500,100,10).
Using these three sets of simulations, we computed the maximum statistics $T_j$, $j=1,2,3$, their confidence intervals and their $p$-value. Table \ref{table:circles} summarizes the results.

\begin{table}[h!]
\begin{center}
\fontsize{8.5}{0.9}\selectfont
\begin{tabular}{llcccc}

statistics&real PD &$(n_b, n_r, n_R)$ & CI & $p$-value& significance\\
\\
\\
\\
\\
$T_1$& 0.2145 &(500,20,50)&[0, 0.1848]&0.0010     &	yes\\
\\
      &       &(500,40,25)&[0, 0.1750]&0.0020     &	yes\\
\\
      &       &(500,100,10)&[0, 0.1713]]&0     &	yes\\
\\
\\
\\
\\
$T_2$& 0.1740 &(500,20,50)&[0, 0.1428]&0.0030     &	yes\\
\\
      &       &(500,40,25)&[0, 0.1402]&0.0010     &	yes\\
\\
      &       &(500,100,10)&[0, 0.1424]&0.0020     &	yes\\
\\
\\
\\
\\
$T_3$& 0.1180 &(500,20,50)&[0, 0.1250]&0.1150     &	no\\
\\
      &       &(500,40,25)&[0, 0.1232]&0.0990     &	no\\
\\
      &       &(500,100,10)&[0, 0.1244]&0.1170     &no\\
\\
\\
\\
\\
\end{tabular}
\end{center}
\caption{{ {\footnotesize  Maximum statistics $T_1$, $T_2$ and $T_3$ for the real persistence diagram and the simulated persistence diagrams of the three circles. The CI is a one-sided confidence interval at a $5\%$ confidence level. The $p$-value is also one-sided. Both the CI and the $p$-value are based on 1,000 simulated persistence diagrams.}}}

\label{table:circles}
\end{table}
The results, for all three scenarios, showed that $T_1$ and $T_2$ were highly significant (the largest $p$-value
reached in any of the six cases was 0.003). In none of the three scenarios was $T_3$ significant, with $p$-values in the range (0.099,\, 0.117). That is, we found that the two points in the $H_0$ persistence diagram (as well as the `point at  infinity', which, recall, we removed from the analysis) are significant. Therefore we have three connected components, as we hoped to find.

%


\end{document}